\documentclass[floatfix,aps,pre,twocolumn,groupedaddress,showpacs,nofootinbib]{revtex4}

\usepackage{amssymb}
\usepackage{graphicx}

\newcommand{\Rm}{{\rm Rm}}
\newcommand{\Rmcrit}{\Rm_*}

\newcommand{\Pm}{{\rm Pm}}
\newcommand{\grad}{\nabla\,}
\newcommand{\Div}{\nabla\cdot}
\newcommand{\curl}{\nabla\times}
\newcommand{\EE}[1]{{\times}10^{#1}}

\begin{document}

\title{Screw dynamo in a time-dependent pipe flow}

\author{Wolfgang Dobler}
\affiliation{Kiepenheuer-Institut f\"ur Sonnenphysik,
  Sch\"oneckstra\ss{}e 6, D-79104 Freiburg, GERMANY}
\email{Wolfgang.Dobler@kis.uni-freiburg.de}

\author{Peter Frick}
\author{Rodion Stepanov}
\affiliation{Institute of Continuous Media Mechanics, 1, Korolev, Perm, 614061,
 RUSSIA}

\date{\today}

\begin{abstract}
  The kinematic dynamo problem is investigated for the flow of a
  conducting fluid in a cylindrical, periodic tube with conducting walls.
  The methods used are
  an eigenvalue analysis of the steady regime, and 
  the three-dimensional solution of the time-dependent
  induction equation.
  The configuration and parameters considered here are close to those of a
  dynamo experiment planned in Perm, which will use a torus-shaped channel.
  We find growth of an initial magnetic field by more than 3 orders
  of magnitude.
  Marked field growth can be obtained if the braking time is
  less than $0.2\,{\rm s}$ and only one diverter is used in the channel.
  The structure of the seed field has a strong impact on the
  field amplification factor.
  Generation properties can be improved by adding ferromagnetic
  particles to the fluid in order to increase its relative permeability,
  but this will not be necessary for the success of the
  dynamo experiment.
  For higher magnetic Reynolds numbers, the nontrivial evolution of
  different magnetic modes limits the value of simple `optimistic' and
  `pessimistic' estimates.
\end{abstract}


\maketitle


\section{Introduction}

The screw dynamo is a simple dynamo model that has been
extensively studied in dynamo theory. It is based on an idea due
to Lortz \cite{Lortz:ExactSol} and Ponomarenko \cite{Ponomarenko:Theory},
according to
which magnetic field can be generated by the helical motion of a
rigid, electrically conducting cylinder of infinite length through
an infinitely extended medium of equal conductivity. This problem
implies a simple velocity field and leads to a critical magnetic
Reynolds number as low as \cite{GailitisFreiberg:TheoryScrewDynamo}
\begin{equation} \label{Rm}
  \Rm = {{U r_0}\over\eta}=17.7 \; ,
\end{equation}
where $U$ is the (constant) longitudinal velocity of the cylinder,
$r_0$ is its radius and $\eta$ is the magnetic diffusivity.
This dynamo model has a
discontinuous velocity profile, and will be referred to as the
`Ponomarenko dynamo'. More realistic models of the screw dynamo,
involving continuous and hydrodynamically realistic velocity
fields were considered by several authors. References
\cite{RuzmaikinEtal:HydroScrew,Gilbert:FastPonomarenko} develop a very
accurate asymptotic theory for the
screw dynamo in smooth flows, which has been complemented by
numerical simulations \cite{Solovyov:ExistenceR}.
In \cite{LupyanShukurov:Screw}, this theory
is applied to a number of realistic flows. Reference
\cite{Leorat:Modelling} extended the numerical analysis to flows
fluctuating in time. Other time dependent screw dynamo models
were presented in \cite{FrickEtal:Nonstationary} in connection with the Perm
dynamo experiment. Reference \cite{Soward:Unified} has put the screw
dynamo into a larger context of slow dynamo mechanisms and
\cite{GilbertPonty:StreamSurfs} have generalized the concept to
non-axisymmetric flows. The nonlinear behavior of the screw
dynamo for spiral Couette flow has been investigated with
asymptotic methods by \cite{BassomGilbert:Nonlinear} and numerically by
\cite{DoblerEtal:NonlinScrew}.

\begin{figure}
  \centering
  \includegraphics[width=0.32\textwidth]{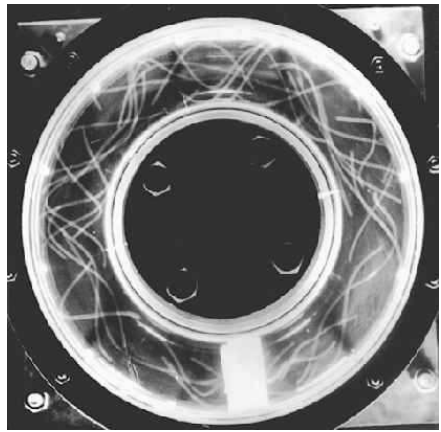}
  \caption{Photograph of a water experiment showing streamlines
    in an initially spinning torus after abrupt braking.
    The white object in the channel is the \emph{diverter}, a kind of a
    fixed ship screw that makes the motion strongly helical.
    From Ref.~\cite{FrickEtal:Nonstationary}.
  }
  \label{Fig-helphoto}
\end{figure}

We are interested in the screw dynamo problem in the context of a
new experimental dynamo project \cite{FrickEtal:Nonstationary}.
The basic idea of this project, introduced in \cite{DenisovEtal:FeasabilityR},
is to realize the dynamo effect in a strongly time-dependent helical flow.
The flow is generated in
a quickly rotating toroidal channel after abrupt braking, and is
shown, for a water experiment, in Fig.~\ref{Fig-helphoto}.
This application raises new questions concerning the screw dynamo,
which have not been addressed in previous studies. In particular,
the flow of the conducting fluid will be located in a {\it
closed} channel and will be supercritical during a {\it short}
time interval only. This requires the solution of the induction
equation for three-dimensional geometry in a time-dependent flow.
Moreover, in oder to understand the saturation of the magnetic
field (if it occurs during the experiment), one has to
investigate the time evolution of the fully nonlinear MHD
equations. The experimental scheme also requires strong
optimization of the channel (minimization of its mass under
optimal conductivity and wall thickness).

In this paper we investigate the screw dynamo in a time-dependent
flow using two different methods which allow different questions
to be addressed. First, the analysis of the eigenvalue problem
related to the case of a steady velocity field gives insight into
the full spectral structure and is numerically the most efficient
approach. Our second method --- numerical solution of the
three-dimensional, space- and time-dependent problem --- is
numerically much more demanding, but it allows us to investigate
the full three-dimensional structure and is the only approach that has
the perspective of tackling the nonlinear problem. While the final
goal is to solve the fully nonlinear problem and to include all
geometrical and dynamical effects, the present paper only
discusses the linear phase of magnetic field growth in (periodic)
cylindrical geometry.

Most laboratory dynamo projects use liquid sodium as conducting
fluid, which has a magnetic Prandtl number
$\Pm = {\nu}/\eta\approx 10^{-5}$.
This means that, in order to achieve the
critical magnetic Reynolds number of a few tens, one operates at
kinematic Reynolds number of order $10^6\text{--}10^7$, which is
far beyond what can be numerically simulated even on the largest
computers. Thus, we cannot solve the dynamical part of the problem
consistently and in the current paper just use mean velocity profiles
as inferred from experiments.

\bigskip

The structure of the paper is as follows.
Section \ref{Formulation} specifies the equations and geometry of the
problem addressed here.
In Sec.~\ref{One-dim}, we present numerical solutions of the induction
equation for prescribed, steady velocity profiles
similar to those found in cylindrical pipes.
If the velocity field is axisymmetric and identical in any cross section
through the cylinder, the problem can be reduced to a one-dimensional
eigenvalue problem that is solved numerically by discretization.
This approach is sufficiently efficient to allow us to scan the
space of relevant parameters and to isolate the cases that will be most
favorable for the realization of the experiment.
Section \ref{Three-d} presents results obtained with a three-dimensional
MHD code solving the induction problem for a velocity field that varies in
space and time.
We finally draw some conclusions about the realizability of the planned
dynamo experiment in Sec.~\ref{Concl}.


\section{Formulation of the problem}
\label{Formulation}

The proposed Perm dynamo experiment \cite{DenisovEtal:FeasabilityR} will
implement a
helical, strongly time-dependent flow of liquid sodium (Na$_{\rm(l)}$) through a
torus, surrounded by a thin shell of copper (Cu), the electrical
conductivity of which is about five times higher than that of
liquid sodium. The role of this conducting shell is to `anchor'
the magnetic field lines in the exterior frame, so that the shear
due to the velocity difference between flow and exterior can
enhance the magnetic field.

Mathematically, dynamo action is characterized by the presence of
growing solutions ${\bf B}({\bf x},t)$ of the induction equation
\begin{equation}
  \frac{\partial{\bf B}}{\partial t}
  = \curl\,[({\bf v}+\eta\grad\ln\mu_{\rm r})\times{\bf B}
    - \eta\curl{\bf B}] \; ,
  \label{Induction}
\end{equation}
satisfying the solenoidality condition
\begin{equation}
  \Div{\bf B} = 0 \; .
  \label{solenoidal}
\end{equation}
Here ${\bf B}$ denotes the magnetic flux density, ${\bf v}$ the
velocity field, $\mu_0$ and $\mu_{\rm r}$ are the magnetic vacuum
permeability and the relative magnetic permeability of the medium,
and $\eta$ is the magnetic diffusivity of the fluid, related to
the electrical conductivity $\sigma$ by $\eta =
(\mu_0\mu_{\rm r}\sigma)^{-1}$.
The term $\eta\grad\ln\mu_{\rm r} \equiv \mathbf{V}_{\rm p}$
arises if the relative magnetic permeability is a function of
position and gives rise to what we call ``paramagnetic pumping'' (with an
effective velocity $\mathbf{V}_{\rm p}$)
of magnetic flux into the regions of enhanced permeability.
A self-consistent description would also include the Navier-Stokes equation
\begin{equation}
  \frac{\partial{\bf v}}{\partial t}
  = -({\bf v}\cdot\grad){\bf v} -\frac{\grad p}{\varrho}
    + \frac{{\bf j}\times{\bf B}}{\varrho}
    + \nu\Delta{\bf v} \; ,
  \label{Navier-Stokes}
\end{equation}
together with the condition
\begin{equation}
  \Div{\bf v} = 0
\end{equation}
for an incompressible fluid.
Here $p$ denotes pressure, $\varrho$ density,
${\bf j} \equiv \curl({\bf B}/\mu_0\mu_{\rm r})$ the electric current
density and $\nu$ is the kinematic viscosity.
The equations given above are complemented by boundary
conditions describing the properties of the walls
(see \cite{DenisovEtal:FeasabilityR}).

In the present paper, we restrict ourselves to the \emph{kinematic
dynamo problem}, i.e., we consider only Eqs.~(\ref{Induction})
and (\ref{solenoidal}), using a given solenoidal velocity field
${\bf v}({\bf x},t)$
and neglect the magnetic back-reaction through the Lorentz force.
This leads to a linear problem in ${\bf B}$ and is very helpful to
understand the evolution for weak and moderately strong magnetic
fields. This approach will allow for optimization of the
experimental apparatus in many respects since the kinematic growth
of ${\bf B}$ over several orders of magnitude is necessary for the
success of the experiment.

The turbulent flow of a conducting medium will give rise to turbulent
induction effects, which can be estimated by mean-field theory.
In the present paper, we neglect these extra terms in the induction
equation (\ref{Induction}) and refer the reader to the discussion in
Ref.~\cite{RaedlerStepanov:Effect}.

While the curvature of the pipe can be expected to play a role for
the geometry of the experiment (the ratio of outer radius $R$ to
inner radius $r_0$ of the torus being about 3), we currently
neglect it by considering a cylindrical pipe instead of a torus.
Connecting the two ends of the cylinder by periodic boundary
conditions and setting its length equal to $L_z=2\pi R$, we obtain
a reasonable first approximation to torus geometry.

\section{One-dimensional, time-independent problem}
\label{One-dim}

\subsection{Equations}

\begin{figure}
  \centering
  \includegraphics[width=0.28\textwidth]{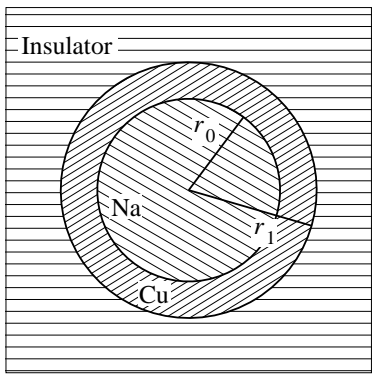}
  \caption{Cross section of the torus/pipe.
    The interior of the pipe, $r<r_0$, is filled with a liquid (sodium) of
    electrical conductivity $\sigma_{\rm fl}$, the solid shell $r_0<r<r_1$
    (made of copper) has a conductivity $\sigma_{\rm sh}$ and is itself
    surrounded by an electrically insulating medium
    (i.e.~$\sigma_{\rm ext}=0$).
  }
  \label{Fig-geometry}
\end{figure}

Let us consider the evolution of the magnetic field in a given
helical flow through a cylindrical, conducting pipe with finite
wall thickness $d=r_1{-}r_0$ (the inner radius of the pipe is
$r_0$, the outer $r_1$), surrounded by an electrical insulator;
Fig.~\ref{Fig-geometry} shows a cross section through the pipe.
We adopt cylindrical coordinates $(r,\varphi,z)$ and assume an
axisymmetric helical velocity field
${\bf v}(r)=[ 0, r \omega(r), v_z(r)]$.
The magnetic diffusivity and permeability may vary as functions of
radius, $\eta = \eta(r)$, $\mu_{\rm r} = \mu_{\rm r}(r)$.

Solutions of the kinematic dynamo problem can be represented as a
superposition of individual modes with exponential growth.
Because the coefficients in the equations depend on $r$ only, we
are looking for solutions in the form of propagating helical
waves
\begin{equation}
  {\bf B}(r, \varphi, z, t) = {\bf b} ({r}) e^{\gamma t +
  i(m\varphi+kz)}\, ,
  \label{Hw}
\end{equation}
where $\gamma$ appears as an eigenvalue and is in general
complex. The real part of $\gamma$ determines whether
${\bf B}$ decays ($\Re\gamma<0$) or grows ($\Re\gamma>0$).
For a given mode, there exists a critical value $\Rmcrit$ of the
magnetic Reynolds number (\ref{Rm}) for which $\Re\gamma$ changes
sign from negative to positive. The lowest value of $\Rmcrit$ is
the threshold for dynamo action.
For the type of dynamo problem
considered here, it is typically between 10 and 100 (see
Fig.~\ref{Fig-gamma-Rm} below for an illustration).

Inserting Eq.~(\ref{Hw}) into the induction equation
(\ref{Induction}), one gets
\begin{eqnarray}
  \lefteqn{ \gamma b_r + i(m \omega+k v_z)b_r+V_{\rm p} \frac{d(rb_r)}{r\,dr} }
    \nonumber \\
  &\qquad=&
  \Rm^{-1}\left[
    \eta\left( \hat{D} b_r - \frac{2 i m}{r^2}b_\varphi \right)
  \right] \; ,
  \label{bsys-r}\\
  \lefteqn{ \gamma b_\varphi
            + i(m \omega+k v_z)b_\varphi+\frac{d(V_{\rm p} b_\varphi)}{dr} }
    \nonumber \\
  &\qquad=&
  r\frac{d\omega}{dr}b_r +
  \Rm^{-1}\Biggl[
    \eta \left(
      \hat{D} b_\varphi + \frac{2 i m}{r^2}b_r
         \right) +
  \label{bsys-phi} \nonumber \\
  && \qquad
  + \frac{d\eta}{dr}
      \left( \frac{d(rb_\varphi)}{r\,dr}
        -\frac{im}{r} b_r
      \right)
  \Biggr] \; ,
\end{eqnarray}
where $V_{\rm p} = \eta\, d\ln\mu_{\rm r}/dr$, and
\begin{equation}
  \hat{D}
  \equiv
  \frac{1}{r} \frac{d}{dr} \left(r\, \frac{d}{dr}\right)
  - \frac{m^2{+}1}{r^2} - k^2
\end{equation}
is a Laplacian-type operator. Equations~(\ref{bsys-r}),
(\ref{bsys-phi}) are written in non-dimensional form: distances
are measured in units of the radius $r_0$, velocity is
measured in units of the longitudinal velocity on the axis of the
flow, $U$. The magnetic diffusivity $\eta(r)$ is measured in units
of $\eta_{\rm fl}$, which introduces the magnetic
Reynolds number
\begin{equation} \label{Rm-Def-2}
  \Rm \equiv \mu_0 \mu_{\rm r,fl} \sigma_{\rm fl} r_0 U
      = \frac{r_0 U}{\eta_{\rm fl}} \; .
\end{equation}
Consequently, in our non-dimensional units we have $\eta=1$
within the fluid and
$\eta=\mu_{\rm r,fl}\sigma_{\rm fl}/\mu_{\rm r,sh}\sigma_{\rm sh}$
in the surrounding shell.

In Eqs.~(\ref{bsys-r}) and (\ref{bsys-phi}), the longitudinal field
component $b_z$ does not enter;
once $b_r$ and $b_\varphi$ are known, it can be derived from the
solenoidality condition $\nabla\cdot{\bf B}=0$, which yields
\begin{equation}
  b_z = \frac{i}{kr} \frac{d}{dr}(rb_r) - \frac{m}{kr} b_\varphi
  \label{hz}
\end{equation}
(note that $k\ne0$ for all growing modes).

The shell $r_0 < r < r_1$ can be treated as a special case of the
above. In fact, in this case $\omega=v_z=0$, and the solution of
(\ref{bsys-r}), (\ref{bsys-phi}) is well-known and given by
\begin{equation}
  b_r \pm i b_\varphi
  = C_\pm I_{m \pm 1}(\kappa r) + D_\pm K_{m \pm 1}(\kappa r) \; ,
\end{equation}
where $\kappa = \sqrt{k^2 + \mu_0 \sigma \gamma}$, and
$C_\pm$ and $D_\pm$ are complex constants determined by the boundary conditions.
Here, $I_m(\cdot)$ and $K_m(\cdot)$
are modified Bessel functions, which are related to the Bessel and Hankel
functions of the first kind by \cite{AbramowitzStegun}
\begin{equation}
  I_m(z) = i^{-m} J_m(iz) \; , \quad
  K_m(z) = \frac{\pi}{2} i^{m+1} H_m^{(1)}(iz) \; .
\end{equation}

A straight-forward approach would be to use this analytical solution and
match it to the solution in the inner region $r < r_0$ (requiring
continuity of the tangential components of the electric field)
and to a potential field in the exterior (see below).
However, we decided to use a simpler approach and solve Eqs~(\ref{bsys-r}),
(\ref{bsys-phi}) in the full region $0 < r < r_1$ for the step-like profile
\begin{equation}
  \eta(r) =
  \left\{
    \begin{array}{l@{\quad}l}
      1                 & ,\ r < r_0 \\
      \sigma_{\rm fl}/\sigma_{\rm sh} & ,\ r_0 < r < r_1
    \end{array}
  \right. \; .
\end{equation}
Here, and for the one-dimensional results presented below, we have set
$\mu_{\rm r}=1$ everywhere.

In the insulating outer domain $r > r_1$, the current density is
zero, $\nabla \times {\bf B}=0$, and thus ${\bf B}$ can be
expressed in terms of a scalar potential $P(r,\varphi,z,t)$,
\begin{equation}
  {\bf B} = -\nabla P \; .
  \label{hp}
\end{equation}
Solenoidality of ${\bf B}$ leads to the potential equation
\begin{equation}
  \frac{1}{r}\frac{\partial}{\partial r}
               \left(r\frac{\partial P}{\partial r}\right)
  +\frac{1}{r^2}\frac{\partial^2 P}{\partial\varphi^2}
  +\frac{\partial^2 P}{\partial z^2}
  = 0 \; ,
  \label{urp}
\end{equation}
and due to the symmetry of the problem $P(r,\varphi,z,t)$  can be written
in the form
\begin{equation}
  P(r, \varphi, z ,t)= p(r) e^{\gamma t+i(m \varphi+k z)} \; .
  \label{pansatz}
\end{equation}
Equations  (\ref{urp}) and (\ref{pansatz}) result in
\begin{equation}
  p''+\frac{1}{r}p'-\left(\frac{m^2}{r^2}+k^2\right)p=0 \; .
  \label{puro}
\end{equation}
Solutions of (\ref{puro}) that are bounded for $r\to\infty$ have the form
\begin{equation}
  p(r)=C K_m(|k|r) \; .
  \label{psol}
\end{equation}

The boundary conditions at $r=r_1$ are obtained from the
requirement of continuity of ${\bf b}$ on the outer border of the
conducting shell. Together with Eqs.~(\ref{Hw}), (\ref{hp}),
(\ref{pansatz}) and (\ref{psol}), this leads to
\begin{eqnarray}
  \frac{b_r(r_1)}{b_\varphi(r_1)}
  &=&
  -i\, \frac{|k|r_1}{m}\,
  \frac{K'_m(|k|r_1)}{K_m(|k|r_1)} \; ,
  \label{gu1} \\
  \frac{b_z(r_1)}{b_\varphi(r_1)}
  &=&
  \frac{k r_1}{m} \; ,
  \label{gu2}
\end{eqnarray}
where $K'_m(x) \equiv dK_m(x)/dx$.
Eliminating $b_z(r_1)$ from Eqs.~(\ref{hz}) and (\ref{gu2}) one finally gets
\begin{equation}
  b_r(r_1)+r_1 b_r'(r_1)
  = -i\left(\frac{k^2r_1^2}{m}+m\right) b_\varphi(r_1) \; .
  \label{gu2n}
\end{equation}
The inner boundary conditions follow from the regularity of ${\bf b}$
at $r=0$ and result in
\begin{equation}
  b_r'(0) = b_\varphi'(0)=0,\qquad  {\rm for} \qquad  \: |m|=1, \atop
  b_r(0) = b_\varphi(0)=0, \qquad  {\rm for} \qquad \: |m|\ne1.
  \label{guc}
\end{equation}

The system (\ref{bsys-r}), (\ref{bsys-phi}), together with the boundary
conditions (\ref{gu1}), (\ref{gu2n}) and (\ref{guc}) is a non-selfadjoint
eigenvalue problem.
Dynamo action implies the existence of eigenvalues $\gamma$ with $\Re\gamma>0$.
To obtain numerically the eigenvalues and eigenfunctions, we replace the
derivatives by their finite-difference counterparts, using 200--800 grid
points for this discretization.
The resulting matrix eigenvalue problem is solved using the
QR-algorithm.

\subsection{Results}

We have checked the one-dimensional numerical code by recalculating growth
rates and
critical magnetic Reynolds numbers from the literature.
In particular, we have considered the case of (infinitely extended)
insulating or perfectly conducting media surrounding the flow (for these
tests we took $d=0.3\,r_0$, and $\sigma_{\rm sh}=0.01\sigma_{\rm fl}$ or
$\sigma_{\rm sh}=100\sigma_{\rm fl}$).
Using the velocity profiles corresponding to the models by \cite{Solovyov:ExistenceR} or
\cite{LupyanShukurov:Screw}, we reproduced the corresponding critical magnetic Reynolds
number with an accuracy of $5\%$ or better.
For the cases with analytical solutions \cite{Ponomarenko:Theory,GailitisFreiberg:TheoryScrewDynamo}, the
accuracy of our numerical results is better than $0.1\%$.

In Ref.~[9], it was demonstrated that the radial profile of the
longitudinal velocity $v_z$ as measured in a water experiment is
reasonably well approximated by
\begin{equation}
  v_z(r) =
  U \frac{\cosh(\xi)-\cosh(\xi r / r_0)}
         {\cosh(\xi)-1}
  \label{hart1}
\end{equation}
for $\xi\approx18$.
Accordingly, we will use this parameterization throughout this paper,
together with
\begin{equation}
  \omega(r) =
  \chi\, \frac{v_z(r)}{r_0} \; ,
  \label{hart2}
\end{equation}
and treat $U$, $\xi$ and $\chi$ as free parameters.
One advantage of the profile (\ref{hart1}) is that
it provides a smooth interpolation between the laminar Poiseuille
solution (for $\xi \to 0$) and rigid-body motion (for $\xi \to \infty$).
The latter limit corresponds to Ponomarenko's model;
in practice, for $\xi=100$, $ d=5\,r_0$ and $\sigma_{\rm sh}=1$,
the critical Reynolds
number differs from Ponomarenko's solution by less than $0.1\%$.

\begin{figure}
  \centering\includegraphics[width=0.35\textwidth]{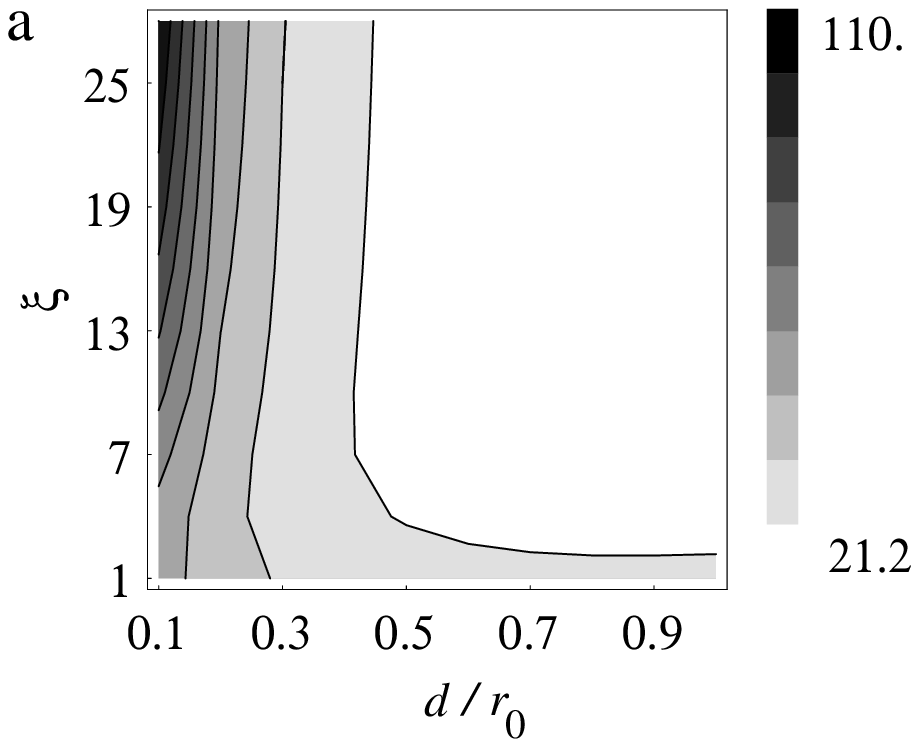}\hfill%
  \centering\includegraphics[width=0.35\textwidth]{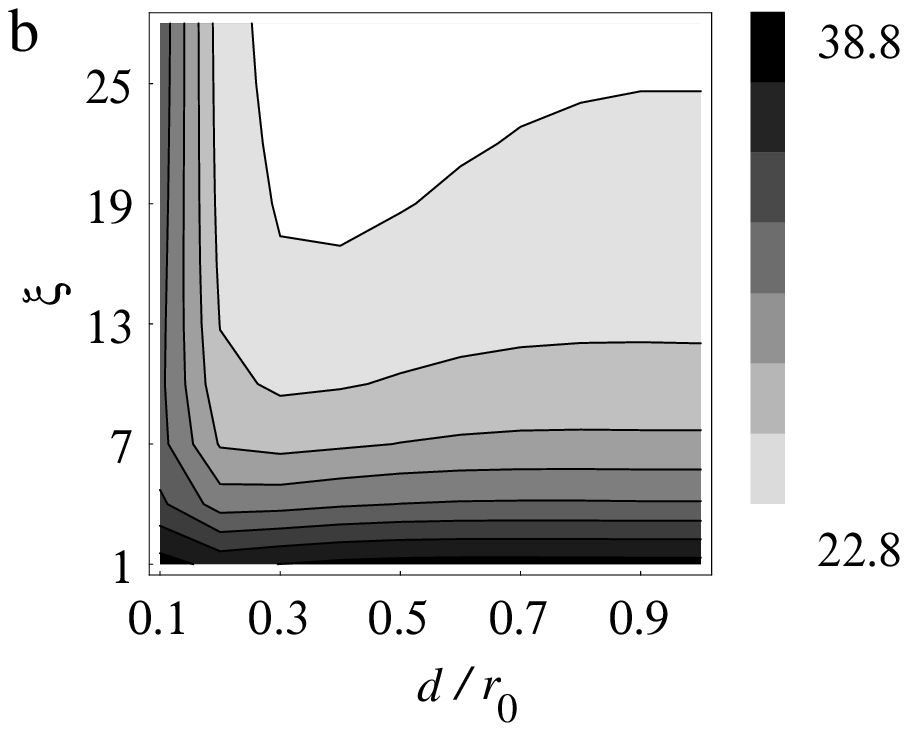}\\
  \caption{Dependence of the critical magnetic Reynolds number $\Rmcrit$ on
    the profile parameter $\xi$ and the shell thickness $d$.
    The graphs show isolines of $\Rmcrit(d,\xi)$ for two conductivity
    ratios:
    a) $\sigma_{\rm sh}=\sigma_{\rm fl}$ and
    b) $\sigma_{\rm sh}=5\sigma_{\rm fl}$.
    All data are for the mode $m=1$, $k=1/r_0$, which is close to the
    fastest growing mode.
  } \label{Rm_xi_d}
\end{figure}

The threshold for dynamo action depends on the conductivity and
the thickness of the shell and on the velocity profile. In
Fig.~\ref{Rm_xi_d} we show the dependence of the critical magnetic
Reynolds number $\Rmcrit$ on the shell thickness $d$ and the
velocity profile parameter $\xi$ for a mode that is close to the
easiest excitable one. In the first case (Fig.~\ref{Rm_xi_d}a) the
conductivities of fluid and shell are equal, in the second case
(Fig.~\ref{Rm_xi_d}b) the shell conductivity is five times higher
than that of the fluid, which approximately corresponds to the combination
Na$_{\rm(l)}$/Cu.

For the case $\sigma_{\rm sh}/\sigma_{\rm fl}=1$ shown in
Fig.~\ref{Rm_xi_d}a, the qualitative dependence of the dynamo threshold on
the profile parameter $\xi$ is different for different values of the wall
thickness.
In the limit of a thin wall ($d \to 0$), $\Rmcrit$ increases monotonously
with $\xi$.
This is explained by the decreasing capability of the shell to `anchor'
magnetic fields lines, which can partially be compensated by a wider shear
region in the fluid, but becomes vital as the flow approaches
rigid-body motion for large $\xi$.
For a thicker shell, we find a very shallow minimum of $\Rmcrit(\xi)$.

For the case of a highly conducting shell, $\sigma_{\rm
sh}=5\sigma_{\rm fl}$, shown in Fig.~\ref{Rm_xi_d}b, the situation
is somewhat different. As $\xi$ increases and thus ${\bf v}$
approaches rigid motion, the threshold decreases, because now
anchoring of the field lines is always given (for the values $d
\ge 0.1\,r_0$ considered here), and the shearing of field lines is
maximized by approaching the discontinuous velocity profile. For
fixed $\xi > 7$, the curve $\Rmcrit(d)$ possesses a minimum at
$d\approx0.3\text{--}0.4$. Thus for the experimentally interesting
values $\xi\approx10\text{--}20$, there is an optimal thickness $d$ of
the conducting shell.



Detailed investigation of the linear dynamo problem
in Ref.~\cite{FrickEtal:Nonstationary} has
shown that in an optimal experimental setup the electric conductivity of
the shell should indeed be approximately five times larger than that of
the liquid sodium.
Provided the shell is thicker than about 15--20\%
of the inner torus radius $r_0$, the actual shell thickness has
little effect on the excitation properties of the magnetic field.
For mechanical reasons, the shell should be kept thin (about
$10\%$ of the inner torus radius), but for the numerical methods
applied in this paper, it is more convenient to consider a
slightly thicker shell (typically about 30\% of the inner radius).

All these results were derived for $\sigma_{\rm ext}=0$, i.e., an
external insulator surrounding the shell, which corresponds well to the
experimental setup.
In the three-dimensional simulations presented in Sec.~\ref{Three-d},
however, we can only approximate such an insulator by setting
$\sigma_{\rm ext}$ to a low but finite value.
A discussion of the error involved by this approximation will be given in
Sec.~\ref{3d-insulator}.


\section{Three-dimensional models}
\label{Three-d}

If the velocity field is time-dependent,
Eqs.~(\ref{Induction}) and(\ref{solenoidal}) can no longer be reduced to an
eigenvalue problem, and we are faced with a Cauchy problem for the
time-evolution of the magnetic field.
Numerical solutions of this problem for the case of $z$-independent
velocity fields have been presented in \cite{FrickEtal:Nonstationary} and
demonstrate that
the screw dynamo should in principle work for the planned dynamo
apparatus, which has an outer (torus) radius $R = 0.4\,{\rm m}$ and
an inner (pipe) radius $r_0 = 0.12\,{\rm m}$.
To make quantitative predictions, however, the longitudinal dependence of
the velocity field needs to be taken into account, since only after a
time comparable to the braking time will all of the fluid be in helical
motion and thus be able to generate magnetic field
(see e.g.~Fig.~\ref{Fig-Uz-t}).
In this Section, we present results for this time- and $z$-dependent flow,
obtained with a three-dimensional MHD code.
We still make the approximation of (periodic) cylindrical geometry and
assume the velocity to be a given function of $r$, $z$ and $t$.
The problem addressed now is thus a three-dimensional kinematic dynamo
problem with space- and time-dependent velocity.

The equation we are solving is the induction equation in the form
\begin{eqnarray}
  \frac{\partial{\bf A}}{\partial t}
  &=&
  {\bf v} \times {\bf B}
  - \eta \mu_0 \mu_{\rm r} {\bf j} +\eta_0\grad\Div{\bf A}
  \nonumber\\
  &=& ({\bf v} + \eta\grad\!\ln\mu_{\rm r}) {\times} {\bf B}
  + \eta \nabla^2\! {\bf A}
       - (\eta{-}\eta_0) \grad\Div\!{\bf A} \quad\quad
  \label{Induction-A}
\end{eqnarray}
for the magnetic vector potential ${\bf A}$, from which the magnetic flux
density ${\bf B}$ and the electric current density ${\bf j}$ are
derived as ${\bf B} = \curl{\bf A}$, and
${\bf j} = \curl({\bf B}/\mu_0\mu_{\rm r})
         = (\curl{\bf B} - \grad\ln\mu_{\rm r}\times{\bf B})/\mu_0\mu_{\rm r}$.
Equation (\ref{Induction-A}) corresponds to the gauge in which the vector
potential ${\bf A}$ and the scalar (electric) potential $\Phi$ are related by
\begin{equation}
  \eta_0 \Div{\bf A} + \Phi = 0 \; ,
\end{equation}
where the constant parameter $\eta_0$ (introduced for purely
numerical reasons) is arbitrary and was chosen equal to the
magnetic diffusivity $\eta_{\rm fl}$ of the fluid.

As before, we use the parameterization (\ref{hart1}), (\ref{hart2}) for
the radial profiles of $v_z$ and $\omega$, where $\chi$ is either 1 (in
Sec.~\ref{3d-steady}) or $z$-dependent and determined from a more
sophisticated model (Sec.~~\ref{3d-timedep}).
All dimensional results in this section refer to the fiducial experimental
apparatus \cite{FrickEtal:Nonstationary} with the parameters given in
Table~\ref{Tab-params}.

\begin{table}
  \centering
  \caption{
    Parameters and results for the different numerical calculations
    presented in Sec.~\ref{Three-d}.
    Parameters common to all models are:
    torus radius $R = 0.4\,{\rm m}$ (resulting in a cylinder length
    $L_z\approx2.5\,{\rm m}$),
    pipe radius $r_0 = 0.12\,{\rm m}$,
    outer shell radius $r_1 = 0.16\,{\rm m}$,
    initial angular velocity of the torus (before braking)
    $\Omega_0 = 310\,{\rm s^{-1}}$,
    and the magnetic diffusivities
    $\eta_{\rm sh}  = 0.016\,{\rm m^2/s}$,
    $\eta_{\rm ext} = 0.4\,{\rm m^2/s}$.
    $T_b$ denotes the braking time.
    The amplification factors $\Gamma_{\rm net}$ and $\Gamma_{\rm max}$
    are defined in Eqs.~(\ref{def-amplif}).
  }
  \label{Tab-params}
  \begin{ruledtabular}
    \begin{tabular}{lcccll}
      Label & $\eta_{\rm fl}$ (${\rm m^2/s}$)
                     & initial field
                                & $T_{\rm b}$ (${\rm s}$)
                                        & $\Gamma_{\rm net}$
                                                      & $\Gamma_{\rm max}$ \\
      \hline
      Run~1  & $0.08$ & random   & $0.1$ & $87$       & $4.4\EE{3}$\\
      Run~1b & $0.08$ & $k=k_3$  & $0.1$ & $1.2\EE{3}$ & $2.1\EE{4}$ \\
      Run~2  & $0.04$ & random   & $0.1$ & $6.8\EE{4}$ & $1.0\EE{6}$\\
      Run~3  & $0.08$ & random   & $0.2$ & $<1$        & $1.9\EE{2}$\\
      Run~4  & $0.04$ & random   & $0.2$ & $860$       & $3.2\EE{4}$\\
  \end{tabular}
\end{ruledtabular}
\end{table}

\begin{figure}
  \centering
  \includegraphics[width=0.4\textwidth]{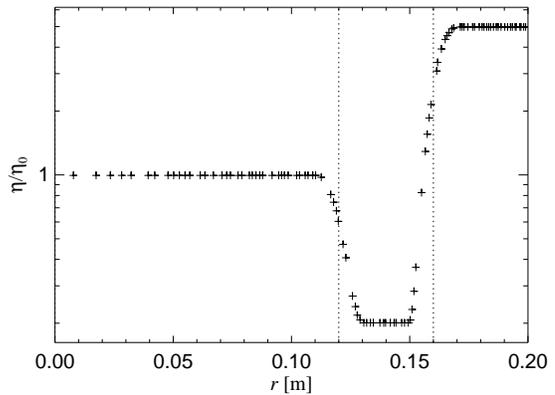}%
  \caption{Radial profile of magnetic diffusivity as used in the numerical
    calculations for the resolution $\delta x = 0.011\,{\rm m}$.
    The region $0 \le r < r_0$ corresponds to the fluid;
    $r_0 < r < r_1$ defines the shell, and $r > r_1$ represents a poorly
    conducting medium surrounding the shell.
    Note that for the tests at $\delta x = 0.0056\,{\rm m}$, the profile
    was steeper and more step-function like.
  }
  \label{Fig-eta-r}
\end{figure}

For numerical reasons we have smoothed the radial profile of magnetic
diffusivity $\eta(r)$; the resulting profile is shown in
Fig.~\ref{Fig-eta-r};
the ratio $\eta_{\rm sh}/\eta_{\rm fl}$ is equal to $0.2$.
We embed the cylinder in a region of enhanced magnetic diffusivity;
while an insulating medium corresponds to $\eta=\infty$ and very large
values of magnetic diffusivity are thus desirable, numerical requirements
limit the values of $\eta$ strongly.
We found that $\eta_{\rm ext} = 5\,\eta_{\rm fl}$ still results in tolerable
time steps, while providing already a good approximation to the case of a
surrounding insulator (cf.~Sec.~\ref{3d-insulator}).

We use a numerical scheme that is of 6th order in space, and perform
3rd-order explicit time-stepping.
Despite the cylindrical geometry, we use a Cartesian grid, which avoids the
special treatment the axis would otherwise require.
The same approach was used with a similar code in
\cite{DoblerEtal:NonlinScrew} to model
nonlinear screw-dynamo action in spiral Couette flow.
Our boundary conditions are periodic in the vertical direction
(corresponding to the model employed in Sec.~\ref{One-dim}).
In the horizontal direction, on the Cartesian boundaries of the
high-diffusivity region, we require the magnetic field to be normal to
the boundaries,
\begin{equation}
  \label{AA-boundconds}
  A_\perp = 0\;, \qquad
  \frac{\partial}{\partial n}{\bf A}_\parallel = {\bf 0} \; ,
\end{equation}
where $\perp$ and $\parallel$ indicate the directions normal and
parallel to the boundary, and
$\partial/\partial n$ denotes the normal derivative.
Conditions~(\ref{AA-boundconds}) imply the absence of currents across the
boundaries which makes it a qualitative local approximation to the case
of a surrounding insulator.

We start with a smoothed random magnetic field.
By this we mean that the vector potential is set to a zero-correlated random
field in all grid points within $r < r_0$ and to zero otherwise;
we then let the magnetic field diffuse  for a fraction $\approx 0.2$ of
the diffusion time $r_0^2/\eta_{\rm fl}$, which reduces the amplitude of
the high-wavenumber modes (which decay quickly anyway) and causes the
field to slightly extend into the shell and, very weakly, into the
external medium.

\bigskip

For the following discussion, it is helpful to characterize the modes by
their longitudinal wave number $k$.
Strictly speaking, this is only appropriate for $z$-independent velocity
fields like the tests given in Sec.~\ref{3d-steady} or towards the end of the
dynamical calculations.
In these cases, individual magnetic modes will have wave numbers
$k_n = n k_1$ which are multiples of $k_1 \equiv 2\pi/L_z$.
However, in many cases one can clearly count the number $n$ of reversals
of $B_z$ along the cylinder, and we will identify this number $n$ with a
corresponding wave number $k_n$.
The azimuthal wave number of any relevant mode is $m=\pm1$
(except for Sec.~\ref{S-mur-2}); for clarity, we will refer only to $m=+1$.

\subsection{Steady velocities}
\label{3d-steady}
To test the code and to get an estimate of the error introduced by the
approximations discussed above (namely the approximation of the
surrounding insulator by a region of enhanced magnetic diffusivity, and the
smoothed radial $\eta$-profile as shown in Fig.~\ref{Fig-eta-r}), we have
compared results of the three-dimensional code with those from the
eigenvalue problem in Sec.~\ref{One-dim} for steady flows.

\begin{figure}
  \centering
  \includegraphics[width=0.45\textwidth]{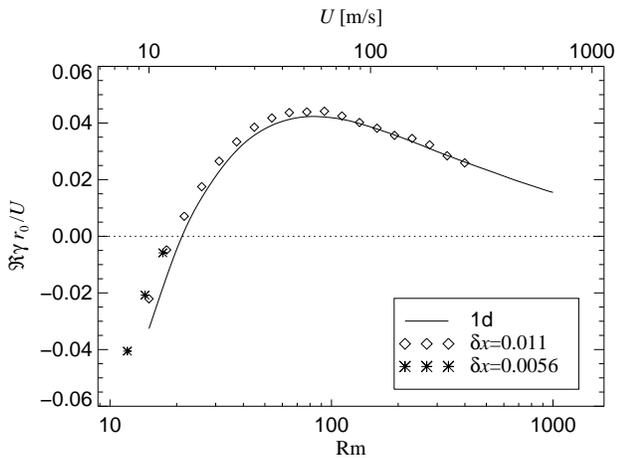}%
  \caption{Kinematic growth rate $\Re\gamma$ as a function of magnetic
    Reynolds number for $d/r_0 = 0.333$.
    Solid line: one-dimensional model with 200 radial points.
    Diamonds ($\diamond$) and asterisks ($*$): values obtained with the
    three-dimensional code at resolution $\delta x=\delta y = 0.011$ and
    $\delta x=\delta y = 0.0056$, respectively.
    The longitudinal wave number is $k=k_3$ in all cases, i.e., the
    longitudinal extent of the pipe is three wavelengths of the magnetic
    mode.
    The second abscissa shows the velocity in an apparatus
    characterized by $r_0 = 0.12\,{\rm m}$, $\eta = 0.08\,{\rm m^2/s}$.
  }
  \label{Fig-gamma-Rm}
\end{figure}

Figure \ref{Fig-gamma-Rm} shows the growth rate $\Re\gamma$ of the magnetic
field obtained by the three-dimensional code (diamonds and asterisks) as a
function of $\Rm$.
The data points are in good agreement with the reference curve, which was
obtained with the one-dimensional code and has an estimated error of
$\le 1\%$.

\begin{figure}
    \includegraphics[width=0.45\textwidth]{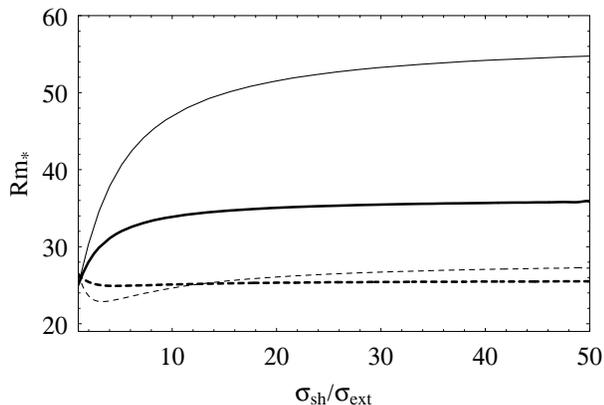}%
  \caption{
    Generation threshold $\Rmcrit$ as a function of the conductivity ratio
    $\sigma_{\rm sh}/\sigma_{\rm ext}$
    as obtained with the one-dimensional model described in
    Sec.~\ref{One-dim}.
    The solid lines are for $\sigma_{\rm sh}=\sigma_{\rm fl}$,
    the dashed line for $\sigma_{\rm sh}=5\sigma_{\rm fl}$.
    Thin lines correspond to a thin shell ($d=0.15\,r_0$),
    thick lines to a thicker shell ($d=0.3\,r_0$).
    All values are for $\xi=10$.}
  \label{Fig-sigma3}
\end{figure}

\label{3d-insulator}
The validity of approximating the insulating exterior region by a
low-conductivity medium can be explicitly assessed from Fig.~\ref{Fig-sigma3},
where the critical magnetic Reynolds number $\Rmcrit$ is shown as a
function of $\sigma_{\rm sh}/\sigma_{\rm ext}$.
An insulator is represented by the limit
$\sigma_{\rm sh}/\sigma_{\rm ext} \to \infty$.
It is evident that $\sigma_{\rm sh}/\sigma_{\rm ext} = 25$ results in
a generation threshold quite close to that for $\sigma_{\rm ext} \to 0$,
and the agreement is particularly good for $\sigma_{\rm sh}=5\sigma_{\rm fl}$.
This is
intuitively clear because the magnetic field is located around the interface
between the liquid and the shell, with low amplitude at the outer shell
surface.

\subsection{Dynamo action in a time-dependent helical flow}
\label{3d-timedep}

Having confirmed the accuracy of our numerical code
for the case of constant velocities, we are now in a position to
investigate the problem for a time- and $z$-dependent velocity field.
We consider only the case of one single diverter in the channel, because
it turns out that additional diverters, while shortening the transition
time after which the flow is fully helical, have a negative impact on the
maximum flow velocity and accelerate the decay of the velocity $U(t)$.
The net effect of increasing the number of diverters has always been found
to be unfavorable for the dynamo.

\begin{figure}
  \centering
  \includegraphics[width=0.47\textwidth]{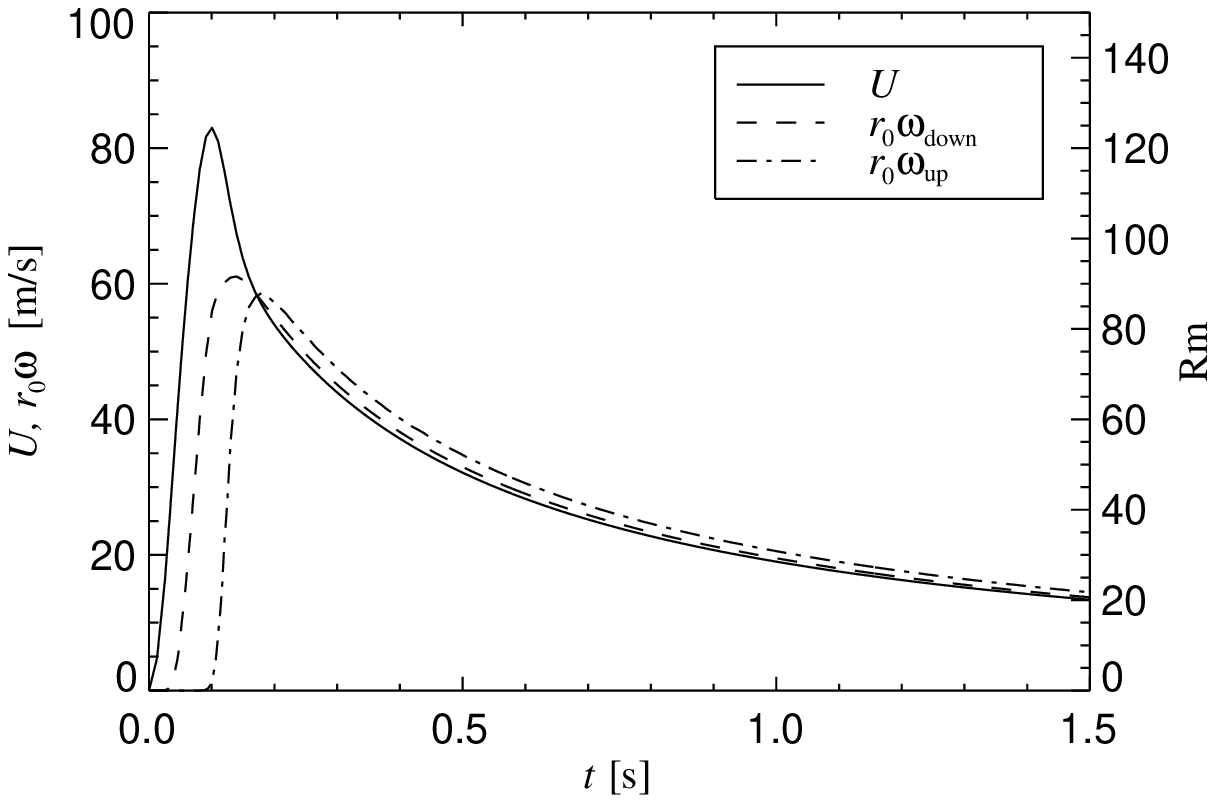}\hfill%
  \includegraphics[width=0.47\textwidth]{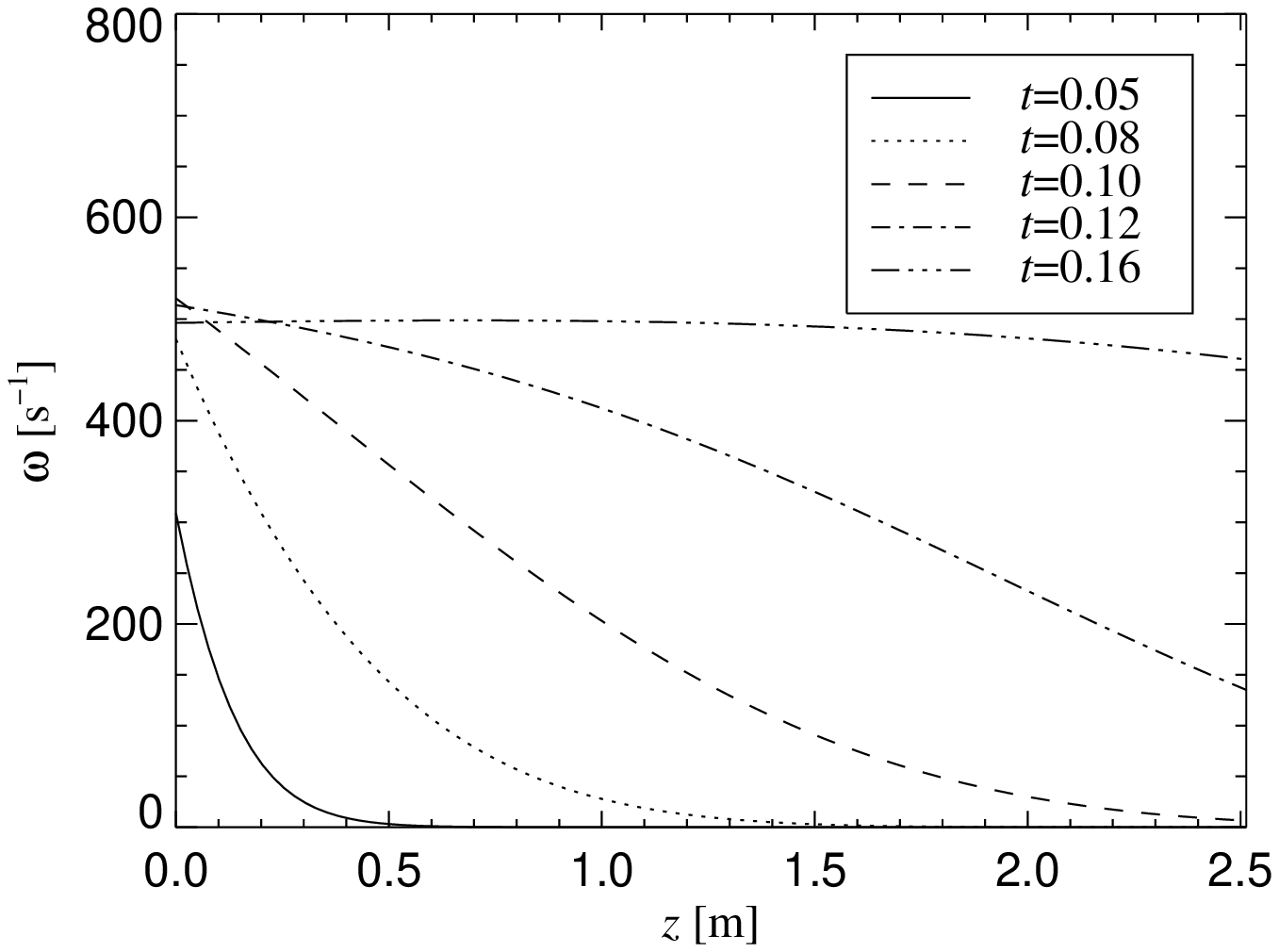}%
  \caption{Evolution of the velocity structure with time.
    The braking time is $T_{\rm b} = 0.1$, and only one diverter is used.
    Top: velocity as a function of time.
    The three curves represent the longitudinal velocity on the axis,
    $U=v_z(r{=}0)$ (solid line), and the angular velocity on the axis
    (multiplied by $r_0$), $r_0\omega(r{=}0)$, downstream and upstream of
    the diverter (dashed and dashed-dotted line, respectively).
    Bottom: Angular velocity $\omega(r{=}0)$ along the cylinder for
    different times.
  }
  \label{Fig-Uz-t}
\end{figure}

We take the longitudinal velocity $v_z(r,t)$ and angular velocity
$\omega(r,z,t)$ from the hydrodynamical model described in
Ref.~\cite{FrickEtal:Nonstationary},
together with the radial dependence (\ref{hart1}), (\ref{hart2}).
At the diverter position, we slightly smooth the velocity field to avoid
discontinuities.
We first consider a short braking time of $T_{\rm b} = 0.1\,{\rm s}$.
Figure~\ref{Fig-Uz-t} shows the time dependence of $v_z$ and $\omega$ on
the axis.
The maximum angular velocity is reached only after the end of the braking
phase and
the rotating zone needs an additional $\approx 0.05\,{\rm s}$ to fill
the whole torus.

\begin{figure}
  \centering
  \includegraphics[width=0.49\textwidth]{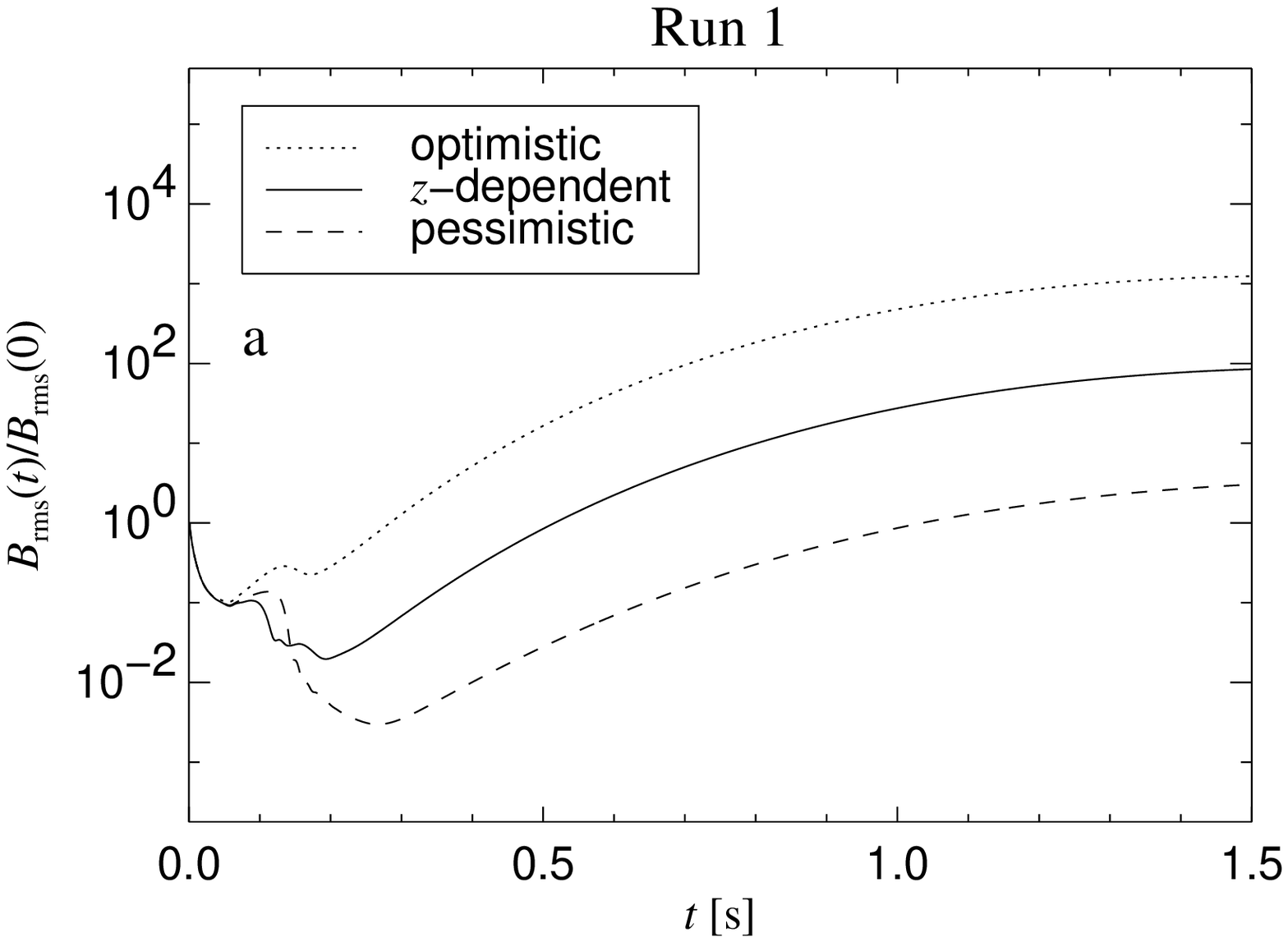}%
  \hfill%
  \includegraphics[width=0.49\textwidth]{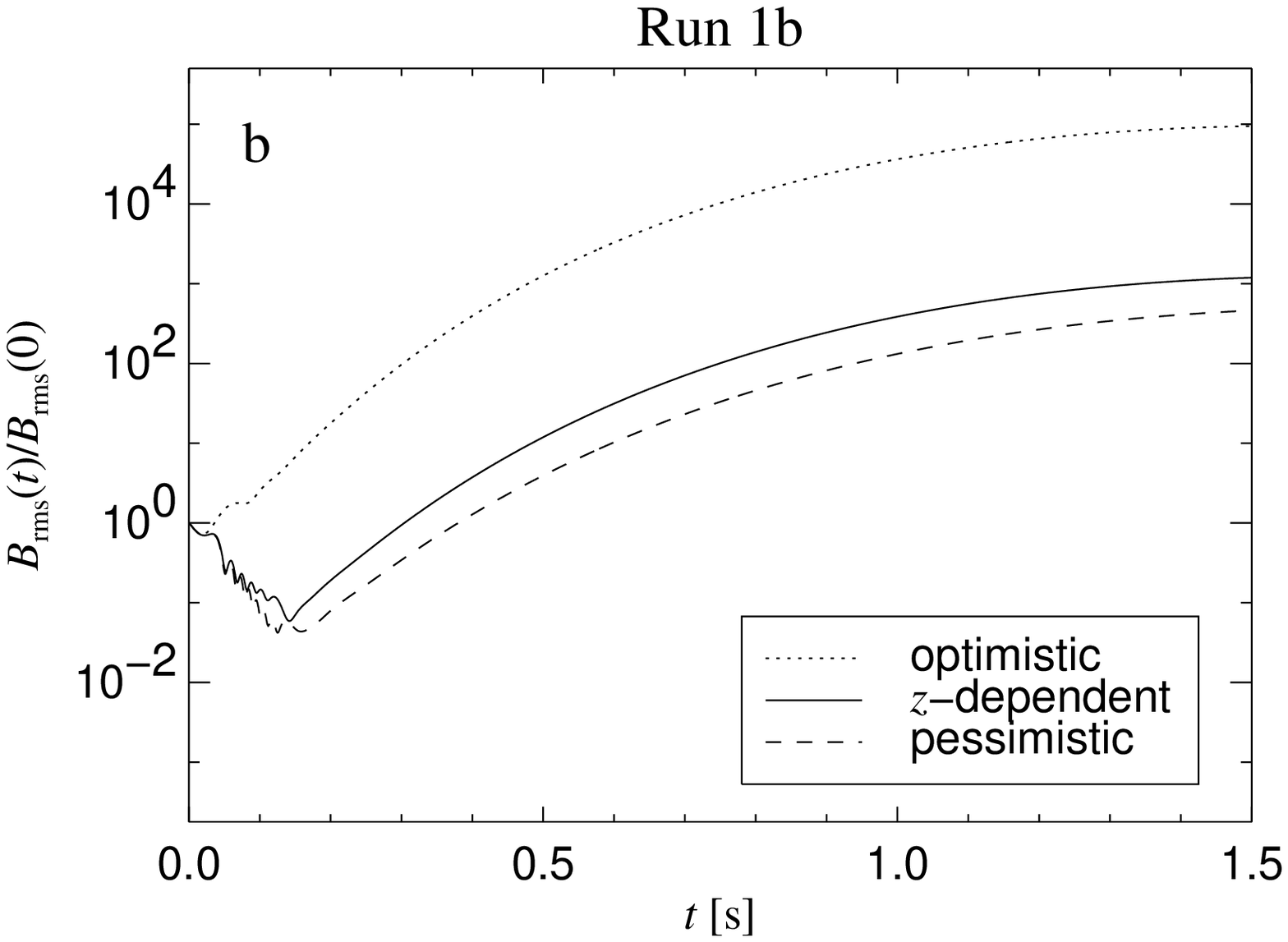}%
  \caption{Root-mean-square magnetic flux density as a function of time
    for three-dimensional models.
    For comparison with the full model (labeled `$z$-dependent'), the
    results of two $z$-independent simulations (`optimistic' and
    `pessimistic' are also shown.
    The spatial resolution here and for all following graphs is
    $\delta x = \delta y = 0.011, \delta z = 0.084$.
    a) Run~1 (starting with a random field).
    b) Run~1b, starting with a clean $k=k_3$ mode.
  }
  \label{Fig-B-t-30I-30M}
\end{figure}

Figure~\ref{Fig-B-t-30I-30M}a shows the time-dependence of the root-mean-square
magnetic field for Run~1, compared to the `optimistic' and `pessimistic'
extrapolations from longitudinally uniform models:
the `pessimistic' model is obtained by adopting the angular velocity
upstream of the diverter for any $z$, while the `optimistic' model
corresponds to using the downstream value \cite{FrickEtal:Nonstationary}.
For $0<t<T_{\rm b}=0.1\,{\rm s}$, the initial field decays towards a
simpler structure (larger scales).
During the time interval $0.1\,{\rm s}<t<0.2\,{\rm s}$, the resulting mode
is restructured into $m=1$, $k=k_3$, which is eventually the fastest
growing mode.
We can define the net and maximum amplification factors as
\begin{equation} \label{def-amplif}
  \Gamma_{\rm net} \equiv \frac{\max\limits_t B_{\rm rms}(t)}
                               {B_{\rm rms}(0)} \; , \qquad
  \Gamma_{\rm max} \equiv \frac{\max\limits_t B_{\rm rms}(t)}
                               {\min\limits_t B_{\rm rms}(t)} \; .
\end{equation}
The corresponding values for Run~1 are $\Gamma_{\rm net}=87$,
$\Gamma_{\rm max}=4.4\EE{3}$ (see Table~\ref{Tab-params}).

\begin{figure*}
  \centering
  \includegraphics[width=0.12\textwidth]{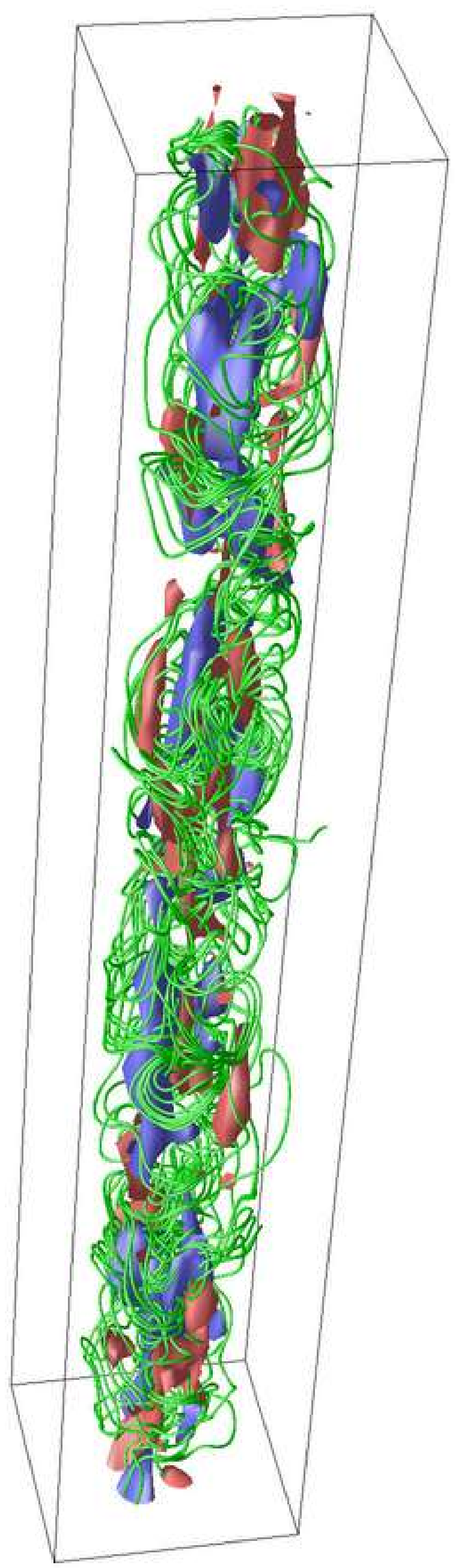}%
  \hfill%
  \includegraphics[width=0.12\textwidth]{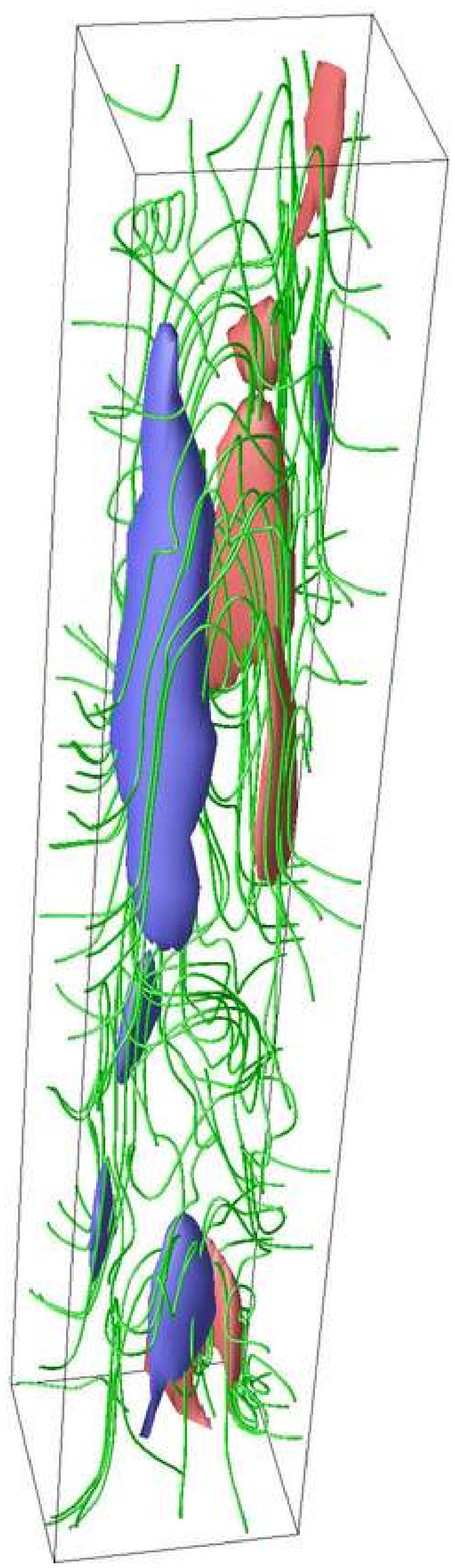}%
  \hfill%
  \includegraphics[width=0.12\textwidth]{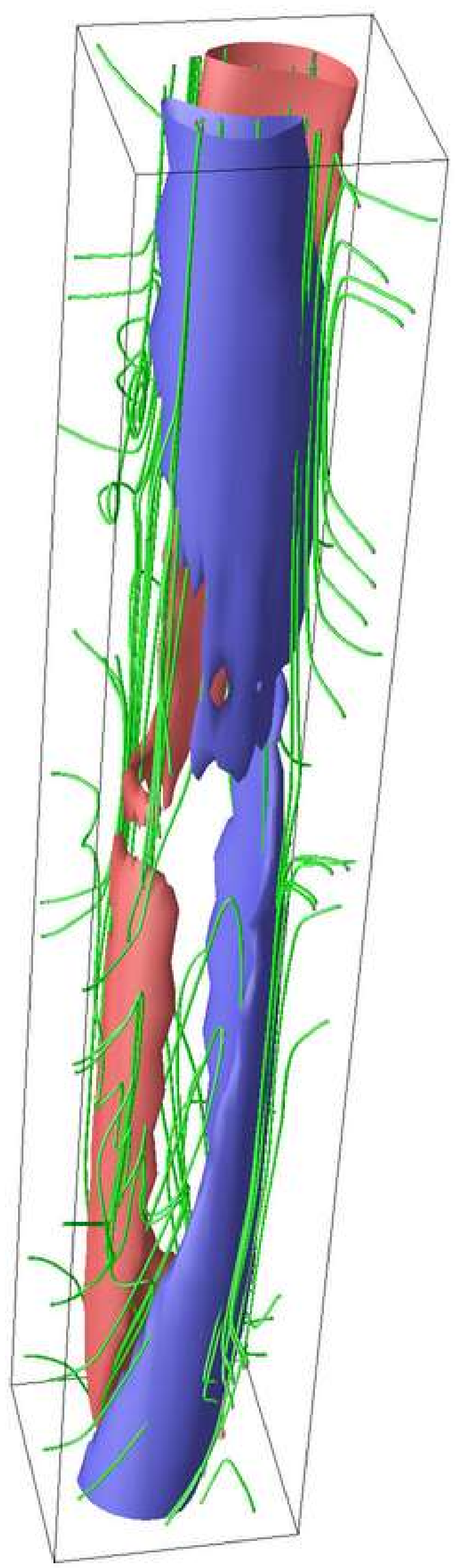}%
  \hfill%
  \includegraphics[width=0.12\textwidth]{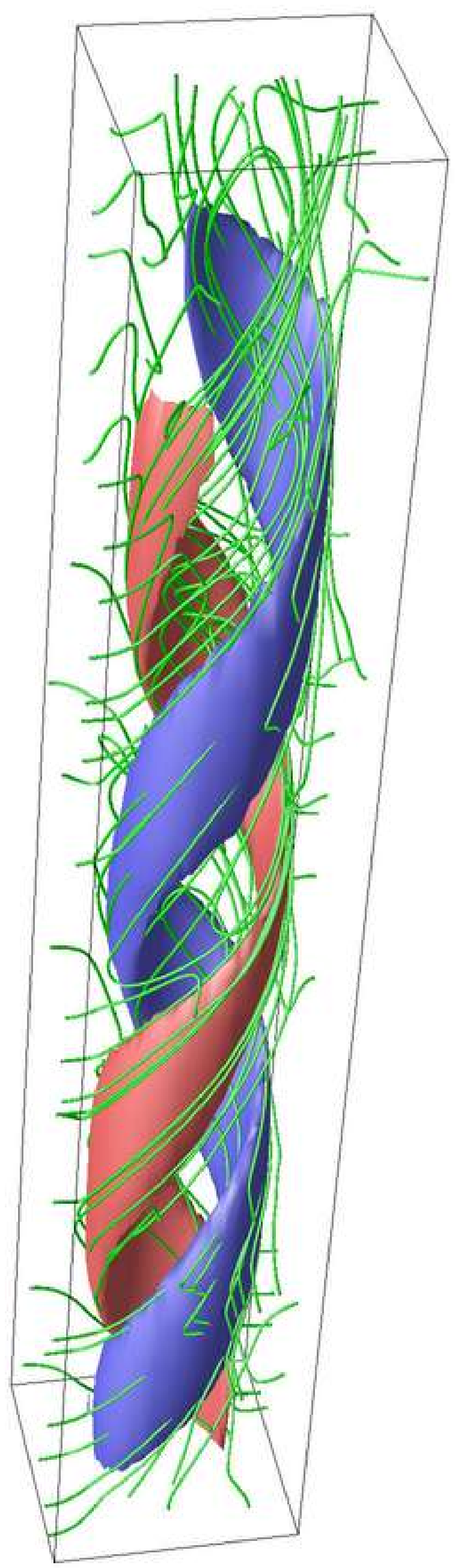}%
  \hfill%
  \includegraphics[width=0.12\textwidth]{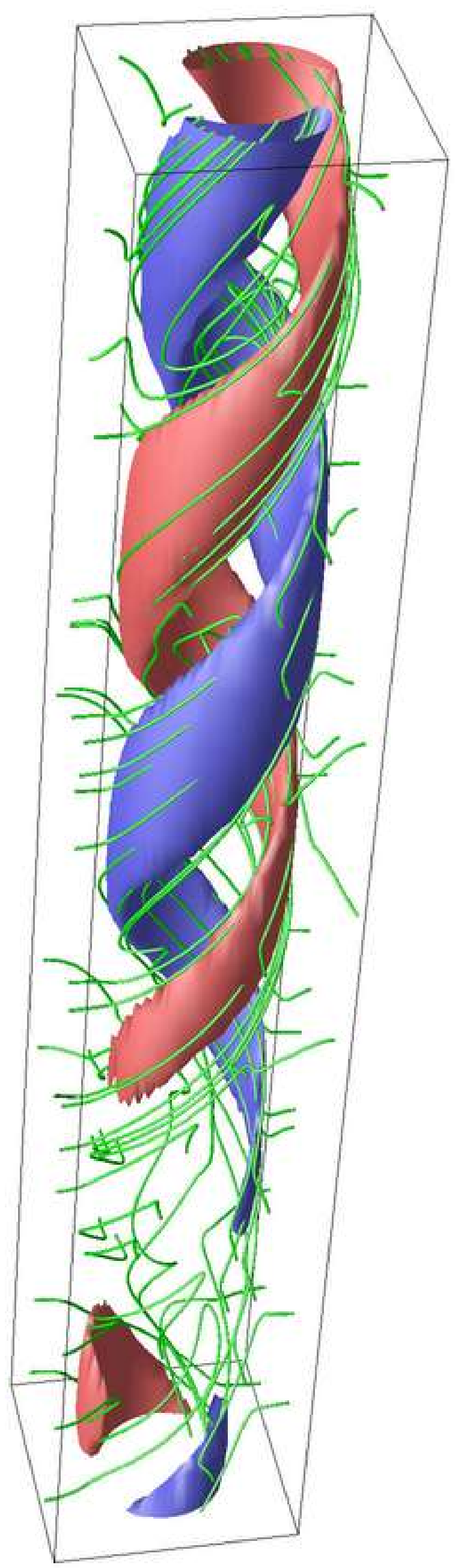}%
  \hfill%
  \includegraphics[width=0.12\textwidth]{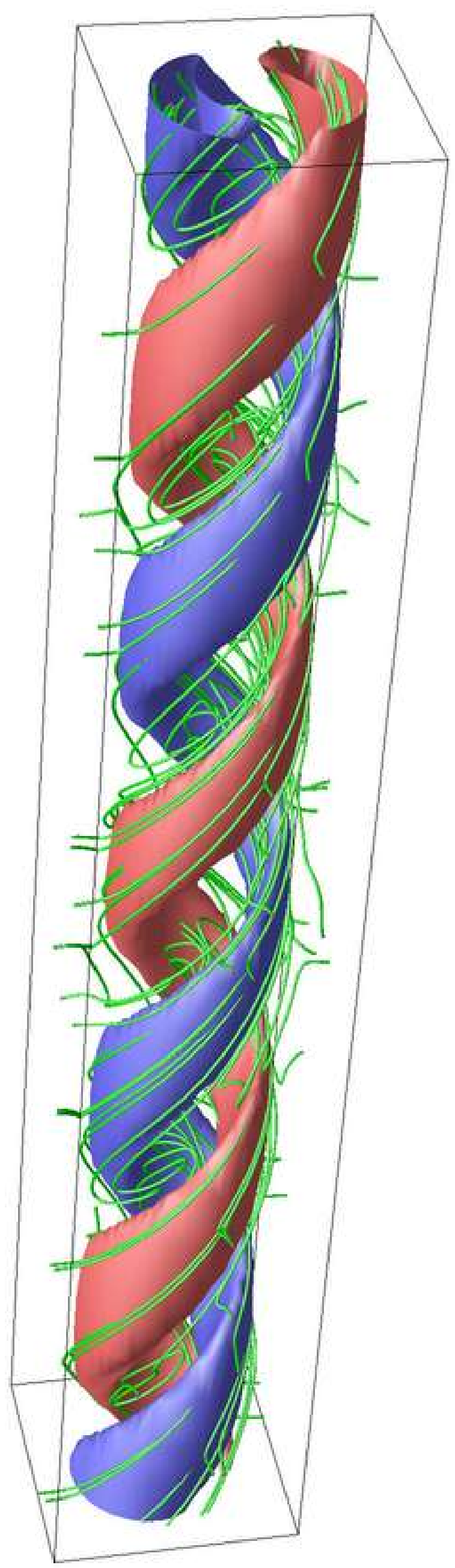}%
  \hfill%
  \includegraphics[width=0.12\textwidth]{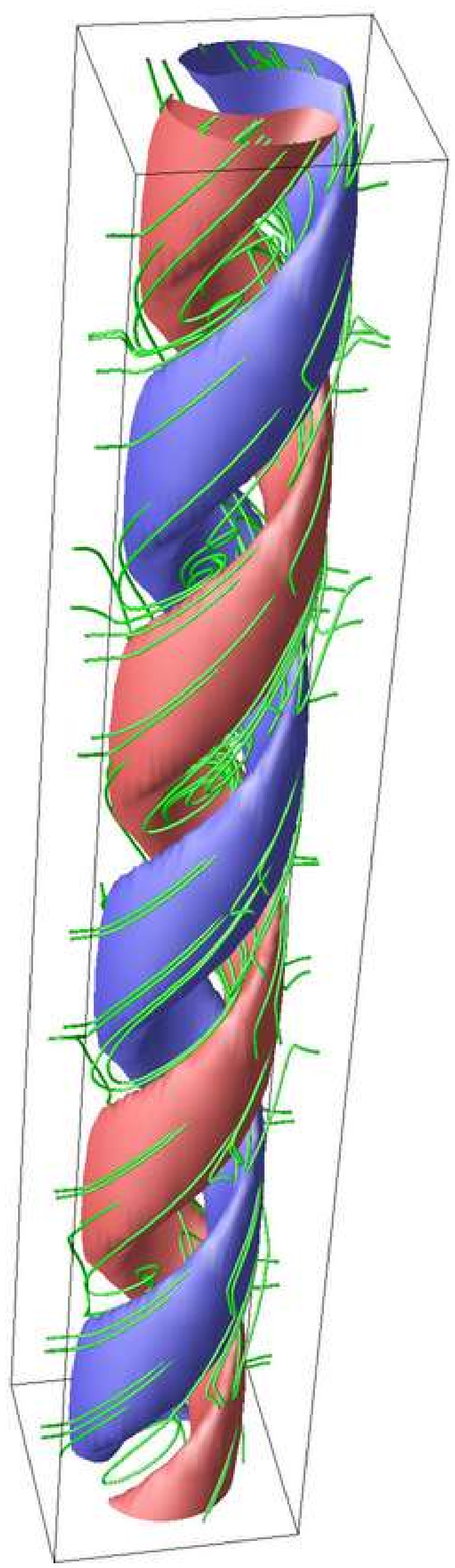}%
  \hfill%
  \includegraphics[width=0.12\textwidth]{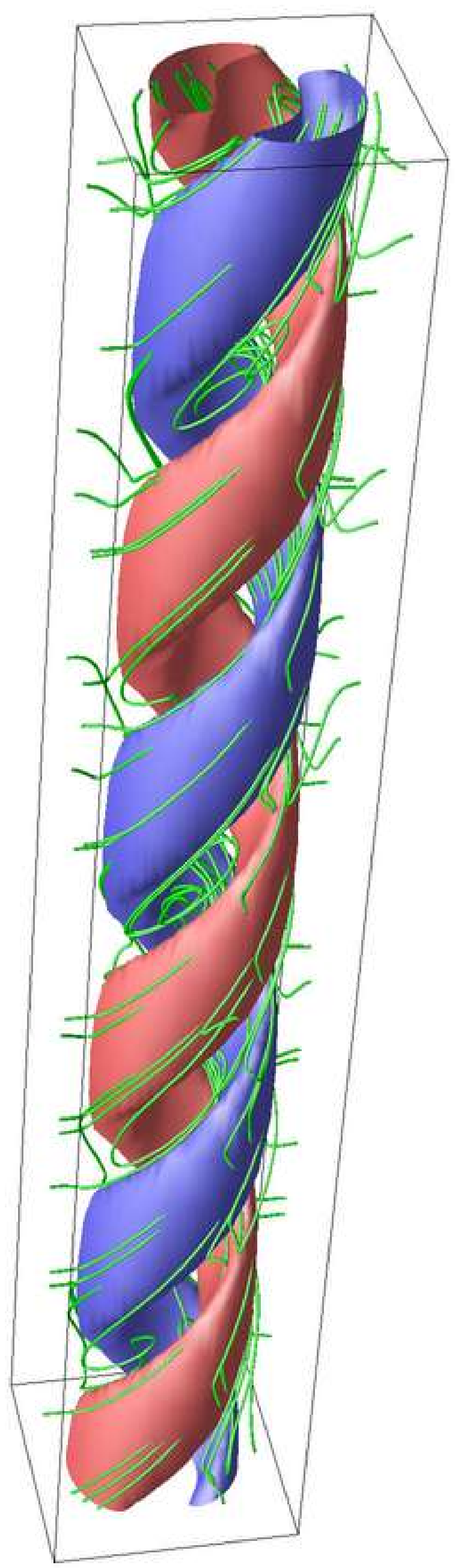}%
  \\
  \caption{Structure of the magnetic field for different times of Run~1.
    From left to right, the times are
    $0.0$, $0.05$, $0.1$, $0.15$, $0.2$, $0.3$, $0.4$ and $0.5\,{\rm s}$;
    the braking time is $T_{\rm b} = 0.1\,{\rm s}$.
    The surfaces are isosurfaces of the magnetic field strength
    (red: $B_z > 0$, blue: $B_z < 0$),
    the lines are magnetic field lines.
    The diverter is located at the bottom and the flow is directed upwards.
  }
  \label{Fig-explorer}
\end{figure*}

The geometrical evolution of the magnetic field structure is
illustrated by Fig.~\ref{Fig-explorer}, where isosurfaces of the
magnetic flux density $|{\bf B}|$ are shown at
eight different times. Note how the initial mode
$m=1$, $k=0$ (the slowest decaying mode with vanishing vertical
net magnetic flux $\Phi_{\rm m} \equiv \int B_z\,dx\,dy$ in a
non-helical flow) is transformed into $k=k_1$ and eventually
$k=k_3$.

Given that modes with different vertical wave numbers $k$
evolve approximately independent from each other
(since the flow always has some $z$-dependence, there is some mixing
between the modes, however), it is not surprising that
the field $k=k_1$ which is
dominant at $t=0.15$ does not provide a good seed field for the
later growth of the final mode $k=k_3$.
This highlights the importance of the initial field configuration for the
net growth of the magnetic field strength during the experiment.
To further investigate the situation, we show in Fig.~\ref{Fig-B-t-30I-30M}b the
growth curve from a different simulation Run~1b, with identical parameters,
but using as initial field the final field (at $t=1.5\,{\rm s}$) from Run~1.
During the first $0.15\,{\rm s}$, the field decays, since the flow is not
helical everywhere.
After $t=0.15\,{\rm s}$, the mode $k=k_3$ grows by a factor of
$2.1\EE{4}$ (maximum growth), and the net growth is $1.2\EE{3}$.
Thus, the choice of the initial field can have a dramatic effect on the
amplification factors.

\subsection{Enhanced magnetic permeability}
\label{S-mur-2}
For a paramagnetic or ferromagnetic fluid, the magnetic diffusivity
$\eta_{\rm fl}$ is
\begin{equation} \label{Rm-murel}
  \eta_{\rm fl} = \frac{1}{\mu_0 \mu_{\rm r,fl} \sigma_{\rm fl}} \; ,
\end{equation}
which shows that one possible option of increasing the magnetic Reynolds
number is to increase the relative magnetic permeability
$\mu_{\rm r}$.
In an experimental setup like that of the Perm
experiment, where there are no movable parts within the fluid,
such an enhancement of $\mu_{\rm r}$ can be safely achieved by
adding ferromagnetic particles to the fluid.
Direct measurements of the effective permeability of two-phase liquids
indicate that $\mu_{\rm r}$ can be at most $\approx 2$ if
reasonable flow properties are to be maintained \cite{FrickEtal:murel}.


\begin{figure}
  \centering
  \includegraphics[width=0.49\textwidth]{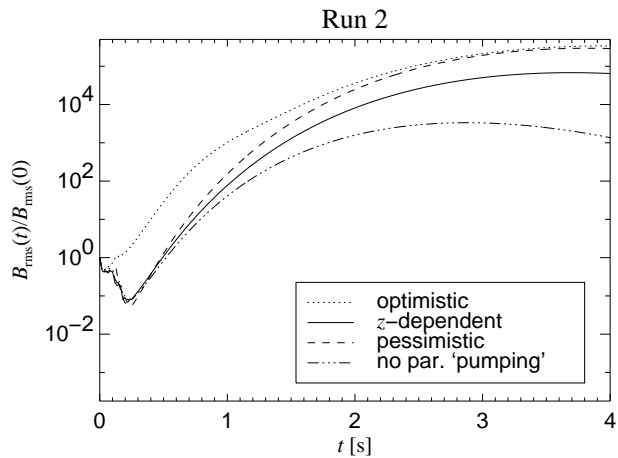}
  \caption{Like Fig.~\ref{Fig-B-t-30I-30M}, but for Run~2, i.e.~for a two
    times lower magnetic diffusivity (achieved by enhancing the magnetic
    permeability to $\mu_{\rm rel} = 2$).
    Note the different time interval plotted and accordingly the much
    longer growth phase compared to Run~1.
    The fourth line (---$\,\cdot\cdot\cdot$) shows the results for a
    model where the `paramagnetic pumping' velocity $\mathbf{V}_{\rm p}$
    was artificially set to zero.
  }
  \label{Fig-B-t-2}
\end{figure}

\begin{figure}
  \centering%
  \includegraphics[width=0.45\textwidth]{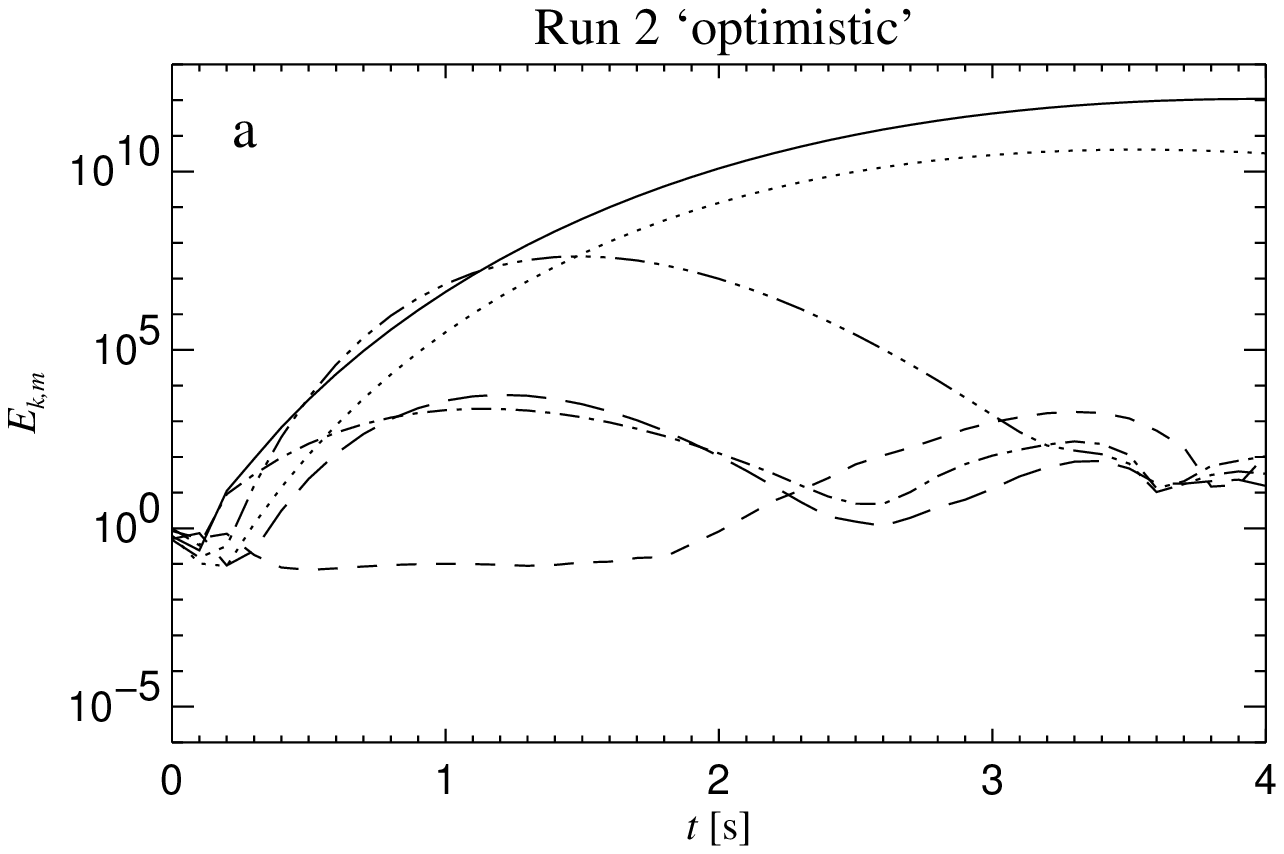}\\%
  \includegraphics[width=0.45\textwidth]{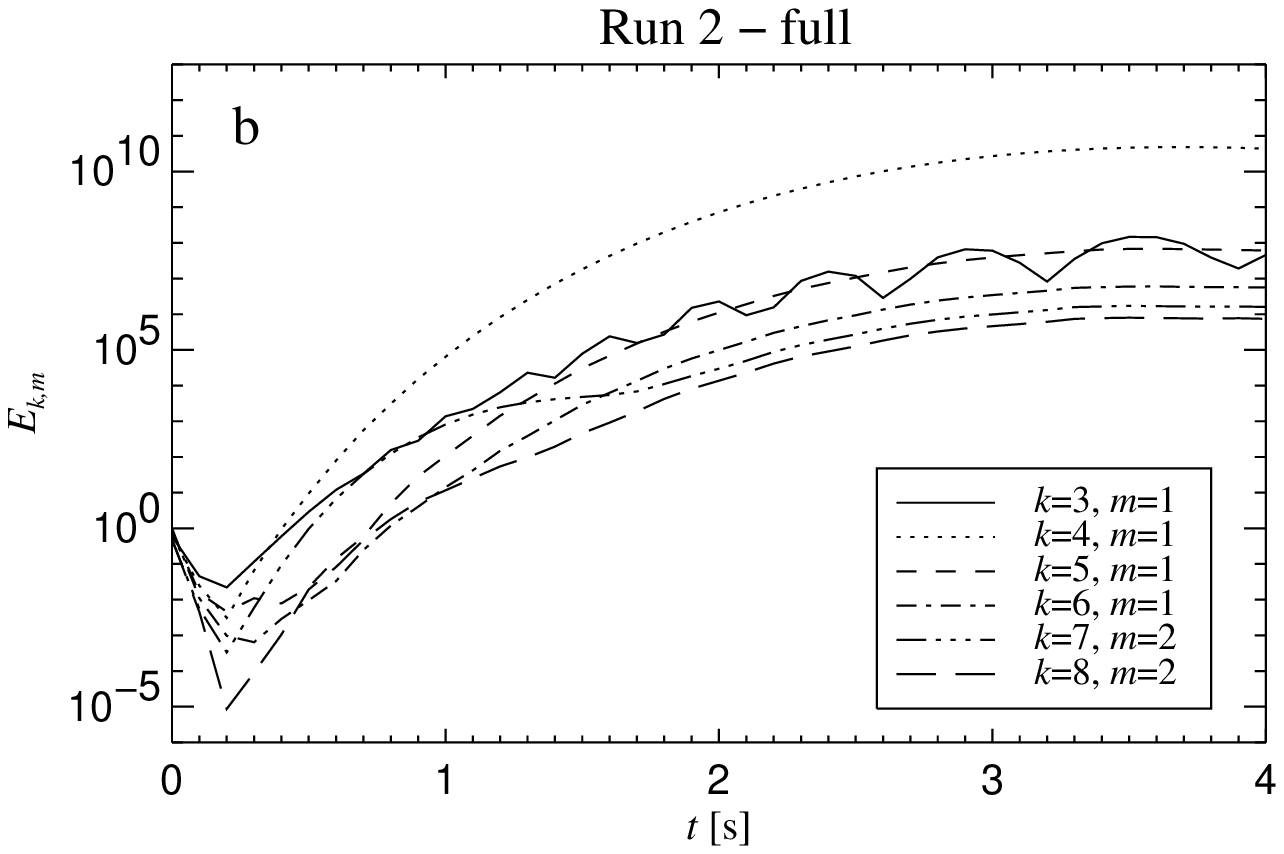}\\%
  \includegraphics[width=0.45\textwidth]{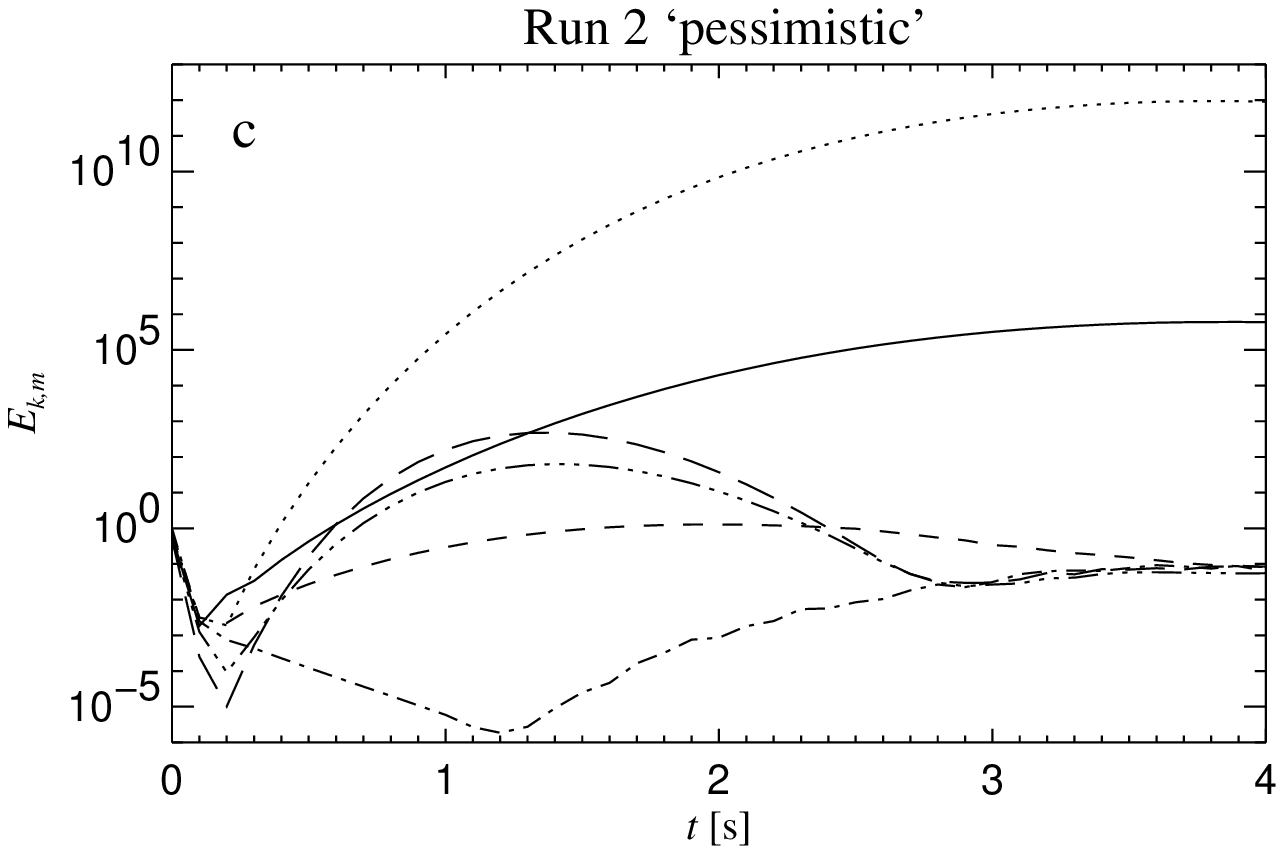}%
  \caption{Evolution of Fourier modes for Run 2.
    a) `optimistic' model;
    b) full model;
    c) `pessimistic' model.
    Shown is the energy $E_{k,m}$ of the modes normalized such that the
    value at $t=0$ is 1.
    The oscillations of the mode $k=k_3$ in the second plot are connected to
    a change in the radial structure, i.e.~most probably due to different
    radial modes.
  }
  \label{Fig-Run2-modes}
\end{figure}

To assess the consequences of such an increase in $\Rm$, we have carried
out as Run~2 a simulation with $\mu_{\rm r,fl}=2$, which implies a two
times lower value of $\eta_{\rm fl}$ (while
$\eta_{\rm sh}$ and $\eta_{\rm ext}$ are fixed to their previous values).
The profile $\mu_{\rm r}(r)$ was a smoothed step profile analogous to the
one shown in Fig.~\ref{Fig-eta-r}, but with only one step at $r=r_0$.
Figure~\ref{Fig-B-t-2} shows that the net and maximum amplification
factors are now increased by about three orders of magnitude.
The main effect is not a faster growth of $B$, but rather a reduced decay
of the initial field, followed by a much prolonged growth phase.
These observations can be understood from Fig.~\ref{Fig-gamma-Rm} as
follows.
Since the dependence $\Re\gamma(\Rm)$ is non-monotonous, doubling the
magnetic Reynolds number will not necessarily increase the growth rate,
but in fact reduce it if $\Rm \gtrsim 40$.
On the other hand, the critical flow speed (corresponding to $\Rm \approx 21$)
is two times lower than for $\mu_{\rm r,fl}=1$, and hence the flow is
supercritical for a much longer time span.

The fourth line in Fig.~\ref{Fig-B-t-2} shows an artificial
$z$-dependent run, where the `paramagnetic pumping' velocity
$\mathbf{V}_{\rm p}$ was set to zero; this would correspond to the case where
$\eta_{\rm fl}$ is reduced by enhancing $\sigma_{\rm fl}$, rather than
$\mu_{\rm r,fl}$.
The comparison shows that the pumping term is indeed important for the
field growth, since without it the net growth would be reduced to about $3400$.

\bigskip

Another interesting finding is that for Run~2 the rms magnetic field for
the full
model is no longer contained in the interval spanned by the `optimistic'
and the `pessimistic' variants.
Rather, the rms field for the `pessimistic' run overtakes the
$z$-dependent one and closely approaches even the `optimistic' run.
This unexpected behavior is connected to the presence of several growing
modes and can be understood by a closer look at the modal structure of the
solutions.
In Fig.~\ref{Fig-Run2-modes} we show the time evolution of individual
modes identified by their vertical wave number $k$.
Strictly speaking, only for $z$-independent velocity profiles (i.e.~for
the `optimistic' and `pessimistic' models) each dynamo mode will be
characterized by a unique value of $k$ (together with the azimuthal wave
number $m$ and a radial one), but spectral analysis for the $z$
direction is a very helpful tool even if this condition is not satisfied.

As can be seen from Fig.~\ref{Fig-Run2-modes}, in the `optimistic' model
the mode $k=k_3$ (and $m=1$) dominates, while for the full and
`pessimistic' models $k=k_4$ (and $m=1$) is the dominating mode.
For most of the time, the growth rate of the modes $k_3$ and $k_4$ are
comparable, which indicates that a wave number between $k_3$ and $k_4$
would be optimal, but is excluded by the geometrical setup.
Other growing modes include $k=k_7$, $m=2$, which was never encountered for
Run~1, where the magnetic Reynolds number was two times lower.

Since the magnetic diffusion time $r_1^2/\eta = 0.36\,{\rm s}$ is at least
comparable to the evolution time of the flow, the growth of the modes is
never just determined by the growth rate for the current value of $\Rm$,
but rather involves the history of the given mode.
For the full, $z$-dependent model, an additional important factor is the
action of the diverter, which for all times (but particularly during the
first $0.2\,{\rm s}$) introduces a $z$-dependence of the velocity field and
thus mixes energy from the dominating mode $k=k_4$ into other modes.
This is the reason why in Fig.~\ref{Fig-Run2-modes}b the different
modes show quite similar behavior after $t\approx 0.3\,{\rm s}$.
Although the same effect will occur for $\mu_{\rm r}=1$, too, the resulting
energy loss from the dominating mode will be weaker there, because the
shorter magnetic diffusion time allows the leading mode to better adjust
to the $z$-dependence of the velocity field.

As a result  of these effects, drawing conclusions for the full problem
from the simple `pessimistic' and `optimistic' models can be quite
problematic.

\subsection{Slower braking of the torus}
\begin{figure}[tbp]
  \centering
  \includegraphics[width=0.49\textwidth]{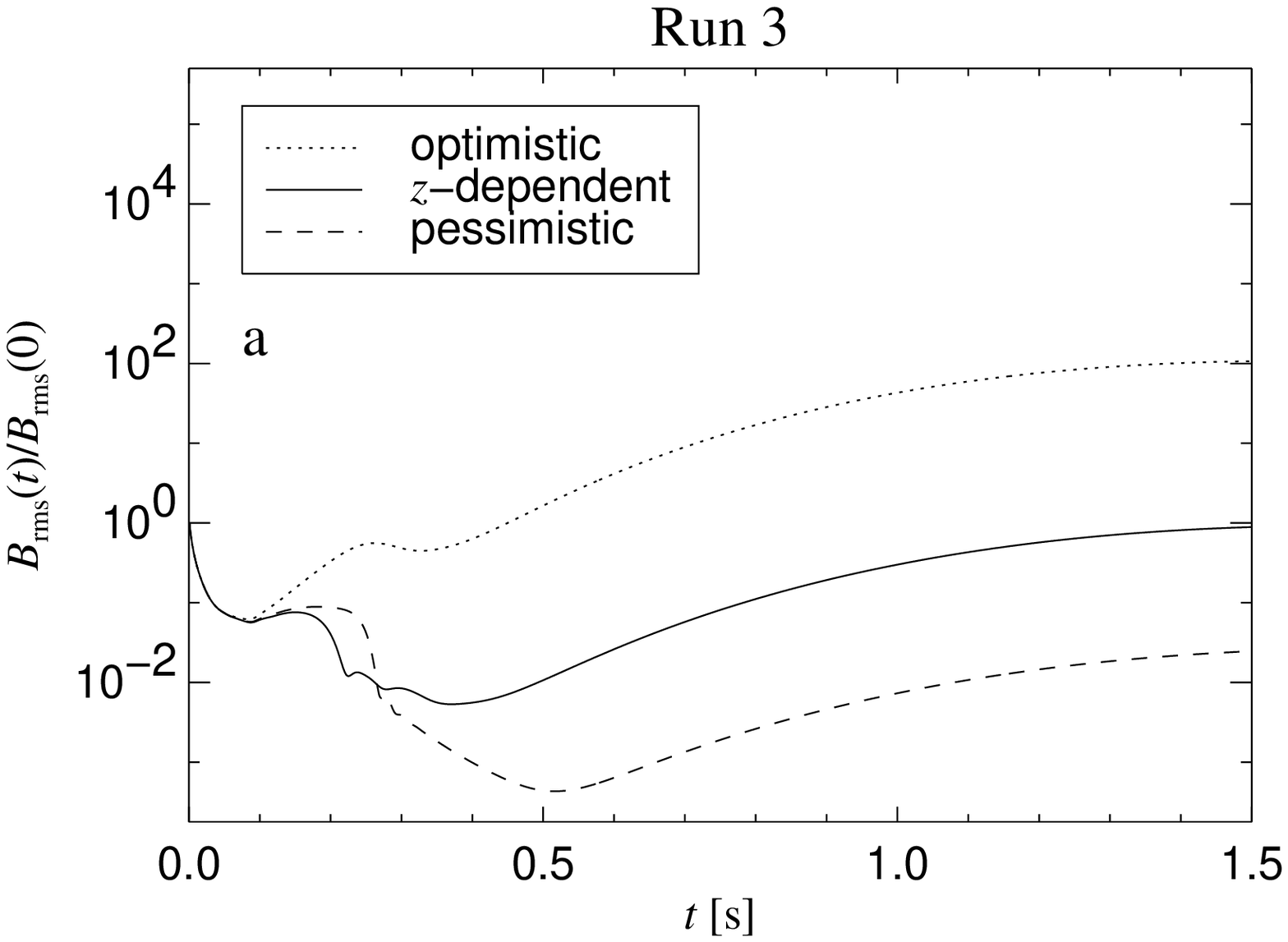}\\%
  \includegraphics[width=0.49\textwidth]{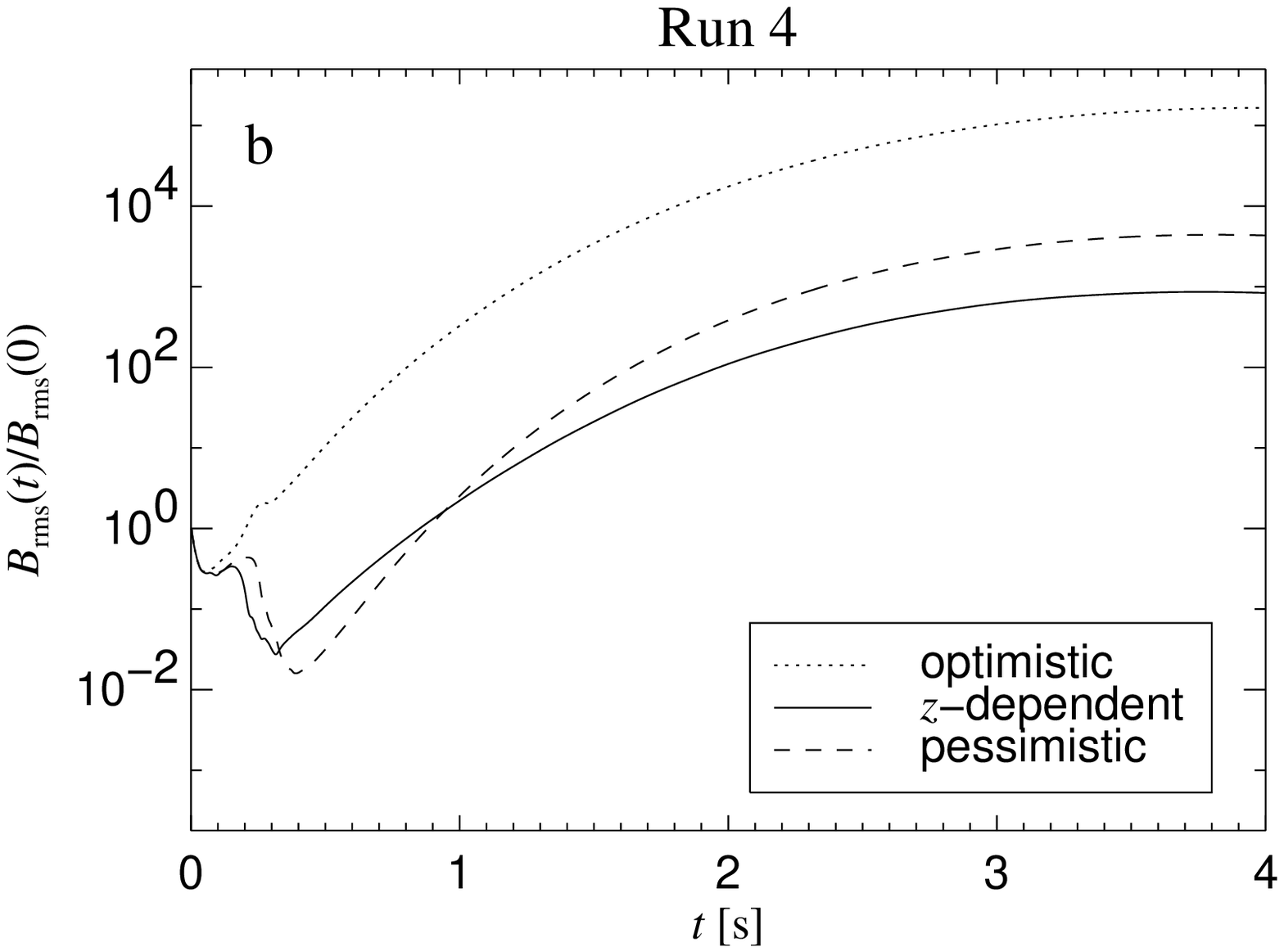}
  \caption{Like Fig.~\ref{Fig-B-t-30I-30M}, but for different parameters.
    a) Like Run~1, but with a longer braking time
    $T_{\rm b}=0.2$ (Run~3).
    b) Like Run~2, but with a longer braking time
    $T_{\rm b}=0.2$ (Run~4).
  }
  \label{Fig-B-t-3-4}
\end{figure}

To assess the effect of a longer braking phase, we have carried out
calculations with $T_{\rm b}=0.2\,{\rm s}$;
the results are shown in Fig.~\ref{Fig-B-t-3-4}a
for $\mu_{\rm r}=1$ (Run~3) and in Fig.~\ref{Fig-B-t-3-4}b for
$\mu_{\rm r}=2$ (Run~4).
In the case of liquid sodium without admixed ferromagnetic particles,
the maximum growth is diminished to about $100$ and the net growth is less
than $1$, i.e., at the end of the experiment the magnetic energy is
lower than it was for the seed field.
Only with the enhanced value $\mu_{\rm r}=2$ we obtain net growth, which is
now larger than in Run~1, but still significantly lower than in Run~2.

\section{Conclusion}
\label{Concl}

The results presented here confirm earlier estimates according to which
the planned Perm
dynamo experiment is realistic and will be able to yield field
amplification factors of about $10^3$ or more.
A thin, highly conducting shell is crucial for the dynamo process and its
role is well understood.

Short braking times are necessary for the dynamo, and a time of
$T_{\rm b}=0.1\,{\rm s}$ as is intended for the Perm experiment
\cite{FrickEtal:Nonstationary} will be sufficient.

Enhancing the magnetic permeability by adding ferromagnetic particles to
the liquid sodium would further enhance the amplification factor, but this
is not crucial for the success of the experiment.
For a longer braking time $T_{\rm b}=0.2\,{\rm s}$, however, enhancing
$\mu_{\rm r}$ is required to obtain net growth of the field at all.
The enhanced magnetic Reynolds numbers for $\mu_{\rm r}=2$ cause a number
of modes to grow (which complicates the analysis), and enhances the
interaction of different modes due to the inhomogeneity introduced by
the diverter.

The final amplification factor strongly depends on the initial magnetic
field configuration, and choosing a suitable seed field can be vital for
obtaining good results.
For example, relying on the terrestrial background field may be
a bad choice, as a uniform field penetrating a torus has only
components $m=0$, $k=\pm k_1$ and $m=1$, $k=0$ while a good seed field
should have a significant amount of energy in the optimal mode $m=1$, $k=k_3$.
A sophisticated arrangement of small permanent magnets may be able to
provide such a field, but one should even consider an arrangement of coils
that allows for a net current through the inner part of the torus.
This question certainly requires more detailed investigations.

\begin{acknowledgments}
  We acknowledge partial financial support from ISTC under grant
  2021, from the Leverhulme Trust (Grant F/125/AL) and from the Deutsche
  Forschungsgemeinschaft (Grant No.~STI 65/12-1).
  RS thanks the Russian Science Support Foundation.
  The three-dimensional numerical calculations have been carried out on
  the PPARC supported parallel computers in St.~Andrews and
  Leicester.
\end{acknowledgments}

\bibliography{dynamo}
\bibliographystyle{apsrev.bst}



\end{document}